\documentclass[10pt,final,journal,twoside]{IEEEtran}%
\usepackage[pdftex]{graphicx}

\usepackage{scalerel}
\usepackage{tikz}
\usetikzlibrary{svg.path}

\definecolor{orcidlogocol}{HTML}{A6CE39}
\tikzset{
  orcidlogo/.pic={
    \fill[orcidlogocol] svg{M256,128c0,70.7-57.3,128-128,128C57.3,256,0,198.7,0,128C0,57.3,57.3,0,128,0C198.7,0,256,57.3,256,128z};
    \fill[white] svg{M86.3,186.2H70.9V79.1h15.4v48.4V186.2z}
                 svg{M108.9,79.1h41.6c39.6,0,57,28.3,57,53.6c0,27.5-21.5,53.6-56.8,53.6h-41.8V79.1z M124.3,172.4h24.5c34.9,0,42.9-26.5,42.9-39.7c0-21.5-13.7-39.7-43.7-39.7h-23.7V172.4z}
                 svg{M88.7,56.8c0,5.5-4.5,10.1-10.1,10.1c-5.6,0-10.1-4.6-10.1-10.1c0-5.6,4.5-10.1,10.1-10.1C84.2,46.7,88.7,51.3,88.7,56.8z};
  }
}

\newcommand\orcidicon[1]{\href{https://orcid.org/#1}{\mbox{\scalerel*{
\begin{tikzpicture}[yscale=-1,transform shape]
\pic{orcidlogo};
\end{tikzpicture}
}{A}}}}

\usepackage{cite}
\ifCLASSOPTIONcaptionsoff
        \usepackage{setspace}
        \usepackage[nomarkers,nolists]{endfloat} %
\fi
\usepackage{amssymb}
\usepackage{amsmath}
\usepackage[T1]{fontenc}
\usepackage[pdfencoding=auto,bookmarks=false,hidelinks]{hyperref} %
\usepackage{algorithmicx}
\usepackage{algpseudocode}
\usepackage{multirow}
\usepackage{siunitx}
\usepackage{booktabs}
\usepackage{pdfrender} %
\usepackage{textcomp} %
\usepackage{wasysym} %
\graphicspath{{./figures/}{./bios/}}

\begin{document}
\bstctlcite{IEEEexample:BSTcontrol} %

\title{Fast and Accurate Retrieval of Methane Concentration from Imaging Spectrometer\\Data Using Sparsity Prior}

\author{Markus~D.~Foote\,\orcidicon{0000-0002-5170-1937},~\IEEEmembership{Graduate Student Member,~IEEE,} 
        Philip~E.~Dennison\,\orcidicon{0000-0002-0241-1917}, 
        Andrew~K.~Thorpe\,\orcidicon{0000-0001-7968-5433}, 
        David~R.~Thompson\,\orcidicon{0000-0003-1100-7550}, 
        Siraput~Jongaramrungruang, 
        Christian~Frankenberg\,\orcidicon{0000-0002-0546-5857}, 
        and Sarang~C.~Joshi%
\thanks{Manuscript received August 6, 2019; revised December 12, 2019 and January 13, 2020 and January 30, 2020; 
accepted January 31, 2020. Date of current version \today.
This work was supported by 
the National Aeronautics and Space Administration (NASA), Grant 80NSSC17K0575. \textit{(Corresponding author: Markus D. Foote)}}%
\thanks{M. D. Foote and S. C. Joshi are with the Scientific Computing and Imaging Institute, 
Department of Biomedical Engineering, University of Utah, 
Salt Lake City, UT 84112 USA (e-mail: foote@sci.utah.edu; sjoshi@sci.utah.edu)}%
\thanks{P. E. Dennison is with the Department of Geography, University of Utah, 
Salt Lake City, UT 84112 USA. (e-mail: dennison@geog.utah.edu)}%
\thanks{A. K. Thorpe, D. R. Thompson, and C. Frankenberg are with the Jet Propulsion Laboratory, 
California Institute of Technology, Pasadena, CA 91109 USA. (e-mail: andrew.k.thorpe@jpl.nasa.gov; david.r.thompson@jpl.nasa.gov; christian.frankenberg@jpl.nasa.gov)}%
\thanks{S. Jongaramrungruang and C. Frankenberg are with the Division of Geological and Planetary 
Sciences, California Institute of Technology, Pasadena, CA 91125 USA. (e-mail: siraput@caltech.edu)}%
\thanks{This is the author's version of the accepted paper. The copy of record may be accessed from IEEE by the DOI \href{https://doi.org/10.1109/TGRS.2020.2976888}{10.1109/TGRS.2020.2976888}\vspace{0.45cm}}%
}%

\markboth{Foote \MakeLowercase{\textit{et al.}}: Fast and Accurate Retrieval of Methane Concentration from Imaging Spectrometer Data Using Sparsity Prior}
{Foote \MakeLowercase{\textit{et al.}}: Fast and Accurate Retrieval of Methane Concentration from Imaging Spectrometer Data Using Sparsity Prior}

\maketitle

\IEEEpubid{\begin{minipage}{1.08\textwidth}\centering\scriptsize\copyright~2020 IEEE. 
        Personal use of this material is permitted. 
        Permission from IEEE must be obtained for all other uses, in any current or 
        future media, including reprinting/republishing this material for advertising 
        or promotional purposes, creating new collective works, for resale or 
        redistribution to servers or lists, or reuse of any copyrighted component 
        of this work in other works.\vspace*{-.2cm}\end{minipage}}

\begin{abstract}
The strong radiative forcing by atmospheric methane has stimulated interest in identifying natural and anthropogenic sources of this potent greenhouse gas. 
Point sources are important targets for quantification, and anthropogenic targets have potential for emissions reduction. 
Methane point source plume detection and concentration retrieval have been previously demonstrated using data from the Airborne Visible InfraRed Imaging Spectrometer Next Generation (AVIRIS-NG).
Current quantitative methods have tradeoffs between computational requirements and retrieval accuracy, creating obstacles for processing real-time data or large datasets from flight campaigns.
We present a new computationally efficient algorithm that applies sparsity and an albedo correction to matched filter retrieval of trace gas concentration-pathlength. 
The new algorithm was tested using AVIRIS-NG data acquired over several point source plumes in Ahmedabad, India. 
The algorithm was validated using simulated AVIRIS-NG data including synthetic plumes of known methane concentration.  
Sparsity and albedo correction together reduced the root mean squared error of retrieved methane concentration-pathlength enhancement by 60.7\% compared with a previous robust matched filter method. 
Background noise was reduced by a factor of 2.64.
The new algorithm was able to process the entire 300 flightline 2016 AVIRIS-NG India campaign in just over 8 hours on a desktop computer with GPU acceleration.

\end{abstract}


\section{Introduction}
\label{sec:intro}

\IEEEPARstart{M}{ethane} (CH$_4$) is a powerful greenhouse gas with a global warming potential 28 times more powerful than CO$_2$ \cite{Saunois2016,Myhre2013}. 
This increased warming potential makes methane responsible for 20\% of the radiative forcing from anthropogenic emissions despite being only 4\% of the carbon mass flux \cite{Kirschke2013,Ciais2013}. 
After a short stable period from 1999 to 2006, atmospheric methane concentration has continued to rise \cite{IPCC2016,Nisbet2014,Nisbet2019}, fueling uncertainty about the partitioning between natural and anthropogenic sources in the global methane budget \cite{Nisbet2016,Saunois2016,Turner2019}. 
Anthropogenic methane sources are dominated by point source emitters in the energy, industrial, agricultural, and waste management sectors \cite{Bogner2003,Beauchemin2005,Allen2013}.
Reduction of anthropogenic methane emissions offers the potential for a rapid reduction in global radiative forcing \cite{Nisbet2019}, but requires identification and mitigation of point source emitters.

\IEEEpubidadjcol

Remote sensing has emerged as a valuable platform for studying methane emissions at various spatial scales \cite{Jacob2016,Duren2016a}.
A variety of satellite instruments for methane measurement have been launched \cite{Frankenberg2011,Strow2003,Yokota2009,Hu2018a}; however, the spatial resolution of these instruments (kilometers to tens of kilometers) is insufficient for studying individual point sources \cite{Ayasse2018}.
Airborne instruments, with a typical spatial resolution less than 10 meters, are well suited for studying individual point sources \cite{Thorpe2014,Thorpe2017}, and the large imaging footprint provided by airborne imaging spectrometers enables the mapping of entire point source methane plumes.
The Airborne Visible/Infrared Imaging Spectrometer Next Generation (AVIRIS-NG) has a demonstrated ability to detect and retrieve concentrations of methane plumes using reflected solar radiance in the shortwave infrared (SWIR; 1400-2500 nm) \cite{Thorpe2013,Thompson2015,Frankenberg2016}.

Methane detections and concentration retrievals using SWIR imaging spectrometer data utilize the strong methane absorption features in this spectral region.
Detection of trace gas plumes from imaging spectrometer data largely builds upon the work of Funk \textit{et al.}\cite{Funk2001}, with subsequent improvements regarding techniques to avoid signal contamination \cite{Theiler2006b,Manolakis2007}. 
Band ratios have also been used to identify plume locations \cite{Bradley2011,Thorpe2013,Thompson2015}.  
Matched filter and band-ratio methods benefit from the ability to sieve through large amounts of data, but they have been inadequate for accurately retrieving methane concentration until the application of a unit absorption spectrum by Thompson \textit{et al.}\cite{Thompson2015}.  
Thompson \textit{et al.} demonstrated real-time concentration retrieval, but this method has not yet been fully validated. 
Iterative maximum a posteriori differential optical absorption spectroscopy (IMAP-DOAS) has also been used to estimate gas concentrations from imaging spectrometer data \cite{Frankenberg2005,Thorpe2014,Thorpe2017,Ayasse2018}. 
However, to date the use of IMAP-DOAS to estimate gas concentration for full scenes and real-time mapping of methane plumes has not been practical because of its high computational requirements. 

In this paper, we present a new algorithm that applies sparsity and albedo correction to retrieval of local enhancements in methane concentration from imaging spectrometer data. 
This algorithm draws from the rich heritage of matched filters for plume detection, and allows accurate retrieval for large datasets from flight campaigns or real-time retrieval. 
We apply this algorithm to real AVIRIS-NG data and to synthetic images for validation.

\section{Methods}
\label{sec:methods}

This section describes the techniques we implement for retrieving methane concentration from SWIR radiance measured by an imaging spectrometer, beginning with previously reported matched filters (Section~\ref{sec:classicmf}) and then reporting our improvements using sparsity (Sec.~\ref{sec:sparse}) and spatial conformity (Sec.~\ref{sec:albedo}). 
Details of the parameters that are used within these techniques are subsequently discussed, including the target spectrum (Sec.~\ref{sec:target}) and background data statistical estimates (Sec.~\ref{sec:estimation}). 
Finally, we describe experimental (Sec.~\ref{sec:data}) and validation (Sec.~\ref{sec:simulation}) datasets used for our evaluation of the reported techniques.

\subsection{Matched Filter}
\label{sec:classicmf}

The radiance measured by the sensor is modeled as a function of the concentration of methane $\alpha$ and a target spectrum $\boldsymbol{t}$ based on the spectral signature of methane absorption $\boldsymbol{s}$, corrupted by additive zero-mean colored Gaussian noise with covariance $\boldsymbol{C}$ capturing the background.
Let $\boldsymbol{L}_0$ be the ambient at-sensor radiance with background concentrations of an absorbing gas, but no enhancement.
The effect of methane enhancement is given by the Beer-Lambert absorption law, $\boldsymbol{L}(\alpha, \boldsymbol{s}) = \boldsymbol{L}_0 e^{-\alpha\boldsymbol{s}}$. 
A quadratic optimization problem is formulated by linearizing the Beer-Lambert absorption law in the Gaussian model using a first-order Taylor series expansion: 
\begin{IEEEeqnarray}{c}
\boldsymbol{L}_0  e^{-\alpha \boldsymbol{s}} 
\approx 
\boldsymbol{L}_0 - \alpha \boldsymbol{t}_{\boldsymbol{s}}(\boldsymbol{L_0}) 
. \IEEEeqnarraynumspace
\end{IEEEeqnarray}
The radiance-dependent target spectrum $\boldsymbol{t}_{\boldsymbol{s}}(\boldsymbol{L_0})$ can be created from the absorption spectrum of methane \cite{Thompson2015}, radiative transfer simulations of transmittance \cite{Thorpe2013}, or radiative transfer simulation of changes in radiance with changes in methane concentration.
We use the latter approach, described in section~\ref{sec:target}.
Following \cite{Funk2001}, we use the mean at-sensor radiance $\boldsymbol{\mu}$ to approximate $\boldsymbol{L}_0$, the unknown nonenhanced ambient at-sensor radiance.
To simplify notation, given that we are interested in a single gas with unchanging characteristic absorption $\boldsymbol{s}$, we drop the subscript in $-\boldsymbol{t}_{\boldsymbol{s}}$ and assume negative values, leaving the target spectrum as simply $\boldsymbol{t}$.
The Gaussian log-likelihood becomes
\begin{IEEEeqnarray}{c}
\hat{\alpha}_i = 
	\underset{\alpha_i}{\operatorname{argmin}}
		\sum_i^N
		\left[ 
			\boldsymbol{d}_{\ref{eq:mf}}^T 
			\boldsymbol{C}^{-1} 
			\boldsymbol{d}_{\ref{eq:mf}}
		\right]  
\IEEEeqnarraynumspace \label{eq:mf}\\*
\boldsymbol{d}_{\ref{eq:mf}} = \boldsymbol{L}_i - ( \boldsymbol{\mu} + \alpha_i  \boldsymbol{t}(\boldsymbol{\mu}) ) \IEEEeqnarraynumspace \nonumber
\end{IEEEeqnarray}
given $\boldsymbol{t}(\boldsymbol{\mu})$, $\boldsymbol{C}$, and $\boldsymbol{\mu}$. 
The minimizer of the above log-likelihood is exactly given by 
\begin{IEEEeqnarray}{c}
\hat{\alpha}_i = 
	\frac{
			\left(
				\boldsymbol{L}_i - \boldsymbol{\mu}
			\right)^T
			\boldsymbol{C}^{-1}
			\left(
				\boldsymbol{t}( \boldsymbol{\mu} )
			\right)
		}{
			\left(
				\boldsymbol{t}( \boldsymbol{\mu} )
			\right)^T
			\boldsymbol{C}^{-1}
			\left(
				\boldsymbol{t}( \boldsymbol{\mu} )
			\right)
		}
.
\IEEEeqnarraynumspace
\label{eq:mfsolution}
\end{IEEEeqnarray}
This minimizer is the same robust matched filter used by \cite{Manolakis2009b} for signature detection in hyperspectral data. 
Strategies for estimating $\boldsymbol{C}$ and $\boldsymbol{\mu}$ are discussed in section~\ref{sec:estimation}.

\subsection{Sparsity Prior}
\label{sec:sparse}

High-dimensional multivariate models are generally difficult to fit reliably due to the large number of potential free parameters.
However, one can recover numerical leverage in the case of sparse data that can be represented by elements with only a handful of nonzero values. 
This property of sparsity is well studied in diverse statistical disciplines, with applications ranging across fields of regression and nonlinear inverse problems.
Sparsity is used in many techniques within statistics, such as the least absolute shrinkage and selection operator (LASSO). 
\textit{Compressed sensing} techniques in image processing use sparse priors to reduce the required measurement time, sampling rate, or consumption of any limited resource during acquisition \cite{Donoho2006}.
In medical imaging, reconstruction methods for magnetic resonance imaging and computed tomography extensively use compressed sensing techniques to reduce scan time and harmful radiation doses to patients \cite{Lustig2007}.
Applications in photography, radio astronomy, and electron microscopy also use compressed sensing \cite{Oike2013a,Wiaux2009a,Stevens2015b}.
Compressed sensing in the imaging sciences is more thoroughly reviewed by \cite{Candes2006}.
The LASSO technique using $\ell_1$ sparsity originated in the geophysics literature \cite{Santosa1986}.
Practical applications of sparsity are particularly focused on $\ell_p$ penalties for $1 \leq p < 2$, because these norms induce sparsity and are convex, giving globally unique solutions. 
In contrast with optimizations using $\ell_2$ priors such as ridge regression, $\ell_1$ regularized optimization problems have no closed-form solution and iterative methods must be employed.

We hypothesize that methane enhancement within airborne imaging spectrometer data should generally exhibit sparsity.
The typical AVIRIS-NG scene has approximately 600 pixels in the cross-track direction and several thousand pixels along-track, totaling to several million pixels. 
Even with multiple point source enhancements that may occupy a few thousand pixels in a scene, these pixels represent a small fraction of the total scene.
We formalize this intuition as a sparse prior on the matched filter optimization.
The number of enhanced pixels within an image is directly measured by the $\ell_0$ counting norm.
The count of pixels containing any enhancement can be included as an additional term in the matched filter optimization problem (\ref{eq:mf}) to yield the minimal number of enhanced pixels.
However, $\ell_0$ is impractical for optimization because it is not convex and requires an exhaustive combinatorial search for the solution \cite{Candes2007}.

Relaxing the $\ell_0$ norm to $\ell_1$ as a convex surrogate norm transforms the problem to a convex optimization with efficient solutions \cite{Optimization}.
The $\ell_1$ norm is the closest convex relaxation of the $\ell_0$ norm.
Furthermore, \cite{Candes2008a} shows that reweighted $\ell_1$ minimization helps rectify $\ell_1$'s dependence on magnitude and tractably solves a relative of the nonconvex $\ell_0$ problem through iterations of convex $\ell_1$ solutions with updating weights.
We introduce this reweighted $\ell_1$ minimization scheme on the same matched filter problem:
\begin{IEEEeqnarray}{c}
\hat{\alpha}^k_i = 
	\underset{\alpha_i}{\operatorname{argmin}}
	\sum_i^N
	\left[ 
		\boldsymbol{d}_{\ref{eq:rwl1}}^T 
		{\boldsymbol{C}^k}^{-1} 
		\boldsymbol{d}_{\ref{eq:rwl1}}
		+ w^k_i 
		\left\| 
			\alpha^k_i
		\right\|_1
	\right]
\IEEEeqnarraynumspace \label{eq:rwl1} \\*
\boldsymbol{d}_{\ref{eq:rwl1}} = \boldsymbol{L}_i - \alpha^k_i  \boldsymbol{t}(\boldsymbol{\mu}^k) - \boldsymbol{\mu}^k \IEEEeqnarraynumspace \nonumber
\end{IEEEeqnarray}
where $\alpha_i$ is the gas enhancement of the $i$-th pixel, $\boldsymbol{C}$ and $\boldsymbol{\mu}$ are the empirical covariance and mean, 
$\boldsymbol{t}(\boldsymbol{\mu})$ is a target radiance spectrum that depends on an approximated ambient at-sensor radiance $\boldsymbol{\mu}$, 
and the regularization weights for the $k$-th iteration are
\begin{IEEEeqnarray}{c}
w_i^k = \frac{
			1
		}{
			\alpha_i^{k-1} + \epsilon
		}
\IEEEeqnarraynumspace
\end{IEEEeqnarray}
where $\epsilon > 0$ is a sufficiently small real scalar for numerical stability when there is no gas enhancement ($\alpha$ is zero). 

The reweighted $\ell_1$ method requires a solution to the $\ell_1$-regularized problem for each iteration of reweighting. 
We use the iterative shrinkage-thresholding algorithm (ISTA) because of its simplicity \cite{Beck2009}. 
The combination of these two iterative methods allows us to alternate iterations and jointly optimize the weights $w_i$ and concentrations $\alpha_i$ for a solution. 
The total computational expense is reduced because an optimal $\ell_1$ solution is not computed at each reweighting iteration. 
ISTA is straightforward to implement, building upon the closed-form solution to the unconstrained optimization problem (\ref{eq:mfsolution}) with only a decrease by the regularization parameter $w$ and subsequent thresholding. 
Gas enhancements are non-negative, so the thresholding is further simplified to allow only positive enhancement values.
The reweighted $\ell_1$ ISTA approach thus iterates independently for each pixel $i$ as
\begin{IEEEeqnarray}{c}
\hat{\alpha}_i^{k} = 
	\max 
	\left(
		\frac{
			\left(
				\boldsymbol{L}_i-\boldsymbol{\mu}^k
			\right)^T 
			{\boldsymbol{C}^k}^{-1} 
			\left(
				\boldsymbol{t}(\boldsymbol{\mu}^k)
			\right)  
			- w_i^k
		}{
			\left(
				\boldsymbol{t}(\boldsymbol{\mu}^k)
			\right)^T 
			{\boldsymbol{C}^k}^{-1} 
			\left(
				\boldsymbol{t}(\boldsymbol{\mu}^k)
			\right)
		}
		, 0
	\right) 
\IEEEeqnarraynumspace
\end{IEEEeqnarray}
where $\boldsymbol{C}^k$ and $\boldsymbol{\mu}^k$ are recalculated for each reweighting iteration.

\subsection{Albedo Correction}
\label{sec:albedo}

Trace gas enhancement is more difficult to detect in pixels acquired over low-albedo surfaces. While trace gas absorption as a percentage of ambient at-sensor radiance is independent of albedo, the absorption signal in terms of absolute radiance is reduced as surface albedo decreases.
In principle, this albedo effect is addressed by construction of the target spectrum, which represents the change in radiance for a unit change in methane enhancement, incorporating surface reflectance.
However, recall that, for simplicity, a single target spectrum is used for many flightlines, and for all the pixels within those flightlines.
This single target spectrum is estimated using the mean of the background distribution. 
A per-pixel estimate of the target spectrum that accounts for local albedo could be more accurate.

We compensate for this nonspecificity effect in our optimization by scaling the target spectrum by a pixel's observed albedo factor, because the absorbed radiance in Beer-Lambert transmission is directly dependent on the initial radiance in the absence of the absorber.
The scalar albedo factor $r_i$ is calculated from the spectral mean $\boldsymbol{\mu}$ and the radiance spectrum $\boldsymbol{L}_i$ of the $i$-th pixel
\begin{IEEEeqnarray}{c}
r_i = 
	\frac{
		\boldsymbol{L}_i^T \boldsymbol{\mu}
	}{
		\boldsymbol{\mu}^T \boldsymbol{\mu}
	}
\end{IEEEeqnarray}
which then scales the target spectrum $\boldsymbol{t}(\boldsymbol{\mu})$. Scaling in this manner results in the solution being normalized by the albedo term. This normalization is similar to per-pixel normalization in the Adaptive Coherence Estimator used in hyperspectral analysis \cite{Manolakis2003,Kraut2005b}.
Additionally, the regularization is scaled by the albedo factor to decrease the regularization of low-signal regions while increasing confidence in retrievals over high-signal regions.
In practice, the scalar albedo factor $r_i$ is factored out of the target spectrum function, yielding the following optimization problem:
\begin{IEEEeqnarray}{c}
\hat{\alpha}^k_i = 
	\underset{\alpha_i}{\operatorname{argmin}}
	\sum_i^N
		\left[ 
			\boldsymbol{d}_{\ref{eq:spatialrwl1}}^T 
			{\boldsymbol{C}^k}^{-1} 
			\boldsymbol{d}_{\ref{eq:spatialrwl1}}
			+ r_i\, w^k_i 
			\left\| 
				\alpha^k_i
			\right\|_1
		\right].
\IEEEeqnarraynumspace \label{eq:spatialrwl1} \\*
\boldsymbol{d}_{\ref{eq:spatialrwl1}} = \boldsymbol{L}_i - r_i \alpha^k_i \boldsymbol{t}(\boldsymbol{\mu}^k) - \boldsymbol{\mu}^k \IEEEeqnarraynumspace \nonumber
\end{IEEEeqnarray}
We again solve this problem with reweighted $\ell_1$ and ISTA. 
The algorithm in \figurename~\ref{alg:spatialrwl1mf} details the procedure for iteratively calculating the optimum gas enhancement solution for the original matched filter, the sparse solution, and this albedo-corrected solution.

\begin{figure}[b!]
	\centering
	\begin{algorithmic}[1]
		\Procedure{AlbedoReWeightL1Filter}{$\boldsymbol{D}$, $\boldsymbol{s}$, $N_{\mathrm{iter}}$}
		\State Initialize $\mu^0 = \frac{1}{N} \sum_{i}^{N} \boldsymbol{L}_i$
		\State Initialize $\boldsymbol{C}^0 = 
								\frac{1}{N} \sum_{i}^{N} 
								\left(
									\boldsymbol{L}_i - \boldsymbol{\mu}^0 
								\right)^T 
								\left(
									\boldsymbol{L}_i   - \boldsymbol{\mu}^0 
								\right)$
		\ForAll{$i$}
			\State Set $r_i = \frac{
									\boldsymbol{L}_i^T \boldsymbol{\mu}
								}{
									\boldsymbol{\mu}^T \boldsymbol{\mu}
								}$ \label{alg:albedostep}
			\State Initialize $ \alpha^0_i = 
									\frac{
										\left(
											\boldsymbol{L}_i - \boldsymbol{\mu}^0
										\right)^T 
										{\boldsymbol{C}^0}^{-1} 
										\left(
											\boldsymbol{\mu}^0 \odot \boldsymbol{s} 
										\right)
									}{
										r_i
										\left(
											\boldsymbol{\mu}^0 \odot \boldsymbol{s}
										\right)^T 
										{\boldsymbol{C}^0}^{-1} 
										\left(
											\boldsymbol{\mu}^0 \odot \boldsymbol{s}
										\right)
									} $ 
			
		\EndFor
		\For{$k = 1$ \textbf{to} $N_{\mathrm{iter}}$} 
			
			\State $\boldsymbol{w}^k = \frac{
											1
										}{
											\boldsymbol{\alpha}^{k-1} + \epsilon
										}$
			\State $\boldsymbol{\mu}^k = 
						\frac{1}{N} \sum_{i}^{N} 
							\left(
								\boldsymbol{L}_i - r_i \alpha^{k-1}_i \boldsymbol{\mu}^{k-1} \odot \boldsymbol{s}   
							\right)$
			\ForAll{$i$}
			\State Let $\boldsymbol{d}_{Ci} = 
				\boldsymbol{L}_i 
				- r_i \alpha^{k-1}_i \boldsymbol{\mu}^{k} \odot \boldsymbol{s}  
				- \boldsymbol{\mu}^k$
			\EndFor
			\State $\boldsymbol{C}^k = 
						\frac{1}{N} \sum_{i}^{N} 
							\boldsymbol{d}_{Ci}^{}\boldsymbol{d}_{Ci}^T 
							$
			\ForAll{$i$}
				\State $\alpha_i^{k} = 
							\max 
								\left(
									\frac{
										\left(
											\boldsymbol{L}_i - \boldsymbol{\mu}^k
										\right)^T 
										{\boldsymbol{C}^k}^{-1} 
										\left(
											\boldsymbol{\mu}^{k} \odot \boldsymbol{s}
										\right) 
										- w_i^k
									}{
										r_i 
										\left(
											\boldsymbol{\mu}^{k} \odot \boldsymbol{s}
										\right)^T 
										{\boldsymbol{C}^k}^{-1} 
										\left(
											\boldsymbol{\mu}^{k} \odot \boldsymbol{s}
										\right)
									}
								, 0
								\right)$
			
			\EndFor
			
		\EndFor
		\State \textbf{return} $\boldsymbol{\alpha}^{N_{\mathrm{iter}}}$
		\EndProcedure
	\end{algorithmic}
	\caption{Algorithm to calculate gas enhancement concentration with albedo-corrected reweighted $\ell_1$ sparsity. 
	For results without sparsity, enforce $\boldsymbol{w}^{k}=0$. 
	For results without albedo correction, set $r_i=1$.}
	\label{alg:spatialrwl1mf}
\end{figure}

\subsection{Target Spectrum Generation}
\label{sec:target}

Matched filter methods use a target spectrum that captures the spectral shape of trace gas absorption. 
We base our target spectrum on radiative transfer simulations, which determine the change in radiance corresponding to a change in methane enhancement above background.
We define a unit absorption spectrum $\boldsymbol{s}$ as the change in radiance for a 1 ppm increase in methane concentration over a pathlength of 1 m.
This spectrum was created using atmospheric radiative transfer simulations in MODTRAN6 \cite{Berk2014}. 
A background methane concentration of 1.85 ppm was assumed, and concentration was uniformly enhanced within a 500 m layer at the surface. 
Accounting for absorption on both the downwelling and upwelling paths, enhancements ranged from 0 to 10000 ppm$\cdot$m. 
Simulations used a 100\% surface albedo, which was necessary to allow $\boldsymbol{s}$ to be rescaled by mean radiance at each wavelength (\ref{eq:elementwisetemplateproduct}).
Sensor height was set to 8 km -- the altitude at which AVIRIS-NG was flown for the test scene described in section~\ref{sec:data}. 
A tropical atmospheric profile with rural aerosol scattering was assumed, and visibility was set to 23 km. 
At-sensor radiance simulated using MODTRAN was convolved to AVIRIS-NG bands using band center wavelengths and full width-half maxima provided with AVIRIS-NG data. 
For each band, concentration multiplied by pathlength (measured in ppm$\cdot$m) was regressed against the natural log of radiance across the range of enhancements, and the slope of the best-fit line provided the unit absorption value for that band (\figurename~\ref{fig:simspectra}d).
This unit absorption spectrum $\boldsymbol{s}$ has unit of $\textrm{(ppm}\!\cdot\!\textrm{m)}^{-1}$.

The target spectrum scales the unit absorption spectrum by the mean radiance at each wavelength:
\begin{IEEEeqnarray}{c}
\boldsymbol{t}_{\boldsymbol{s}} (\boldsymbol{\mu}) =\boldsymbol{\mu} \odot  \boldsymbol{s} \label{eq:elementwisetemplateproduct}
\end{IEEEeqnarray}
where $\odot$ is element-wise multiplication of the vectors.
As $\boldsymbol{\mu}$ has units of $\mu \textrm{W}\! \cdot\! \textrm{cm}^{-2}\!\cdot\! \textrm{sr}^{-1}\!\cdot\!\textrm{nm}^{-1}$ for AVIRIS-NG data, the target spectrum has units $\mu \textrm{W}\! \cdot\! \textrm{cm}^{-2}\!\cdot\! \textrm{sr}^{-1}\!\cdot\!\textrm{nm}^{-1} \textrm{(ppm}\!\cdot\!\textrm{m)}^{-1}$. 
Since the target spectrum is a scaled change in radiance due to a 1 ppm increase in concentration over a pathlength of 1 m, $\alpha_i$ is the enhancement above the background of a combined \textit{concentration-pathlength} with units of ppm$\cdot$m.
This measurement unit reflects an inherent ambiguity in radiative retrieval: a low-concentration enhancement that persists over a long pathlength is radiatively equivalent to a high enhancement over a short pathlength.

\subsection{Strategies for Estimating \texorpdfstring{$\boldsymbol{C}$}{C} and \texorpdfstring{$\boldsymbol{\mu}$}{\textmu}}
\label{sec:estimation}

In the matched filter model, the Gaussian modeling parameters $\boldsymbol{C}$ and $\boldsymbol{\mu}$ describe the \textit{background} signal of the data. 
Any contamination of the target spectrum into these parameters, especially contamination in the covariance matrix~$\boldsymbol{C}$, degrades the detection ability of a matched filter \cite{Theiler2006b}. 
Various techniques have been reported in the literature for covariance estimation and inversion \cite{Manolakis2009b}. 

For the robust matched filter method, we estimate the covariance with a robust approach described by Theiler~\cite{Theiler2012} with a mean Mahalanobis approximation. 
This robust covariance estimation requires an \textit{exhaustive search} through many proposed shrinkage values for the one that best minimizes the average leave-one-out negative log likelihood. 
For our iterative estimations, we use an iterative approach that allows us to directly remove any gas enhancement signals from these values. 
An iterative approach is less computationally demanding than an exhaustive search and ensures that the background covariance is not contaminated by the target. 
The mean and covariance are calculated with their canonical formulas after subtracting the current signal estimate from the data. 
First, the mean is calculated from the data with the removal of the most recent enhancement estimates $\alpha_i^{k-1}$:
\begin{IEEEeqnarray}{c}
\boldsymbol{\mu}^k = 
	\frac{1}{N} \sum_{i}^{N} 
		\left(
			\boldsymbol{L}_i 
			- \alpha^{k-1}_i  \boldsymbol{t}(\boldsymbol{\mu}^{k-1}) 
		\right). \IEEEeqnarraynumspace
\end{IEEEeqnarray}
The covariance is then calculated with updated mean $\boldsymbol{\mu}^k$ and the most recent enhancement estimates $\alpha_i^{k-1}$:
\begin{IEEEeqnarray}{c}
\boldsymbol{C}^k = 
	\frac{1}{N} \sum_{i}^{N} \boldsymbol{d}_{\ref{eq:cov}}^{}\boldsymbol{d}_{\ref{eq:cov}}^T.
\IEEEeqnarraynumspace \label{eq:cov} \\*
\boldsymbol{d}_{\ref{eq:cov}}^{} = \boldsymbol{L}_i 
- \alpha^{k-1}_i  \boldsymbol{t}(\boldsymbol{\mu}^k) 
- \boldsymbol{\mu}^k 
\IEEEeqnarraynumspace \nonumber
\end{IEEEeqnarray}

Due to the slightly nonuniform response of individual detectors in pushbroom instruments, models applied to data from pushbroom instruments have estimated the covariance independently for each detector \cite{Thompson2015, Thompson2016}.
Practically, each flightline image is partitioned by the detector element that collected the image pixel's data, and each partition is processed independently. 
This detector correspondence is preserved along the image columns (direction of flight) in the non-orthorectified data. 
Full partitioning by the detector naturally leads to more computational work, because covariance and mean calculations are repeated within the independent matched filter processes for each individual detector-wise partition of the image data. 
To ease this computational load, we introduce a collection of multiple adjacent detectors' pixels into a single partition for processing.
This approach is especially useful for short flightlines where the limited sample size leads to a less accurate background model. 
The number of adjacent detectors grouped into a single processing partition is a trade-off among the sample size for estimating the background model, the background model's accurate description of individual detector variations, and the reduced computational complexity of having fewer partitions.
In our experiments, this detector grouping is applied across five adjacent detectors for unmodified instrument data. Detector grouping is not applied in experiments using simulated data for better quantitative comparison to previous work. For the convergence experiment, we group all detectors into one partition for the simplicity of having a single optimization energy.

\subsection{Experimental Data}
\label{sec:data}

The NASA AVIRIS-NG sensor is a pushbroom-design imaging spectrometer operated by the Jet Propulsion Laboratory (JPL) covering a spectral range of 380-2510 nm with bands centered at approximately 5 nm intervals \cite{Hamlin2011}. 
For this study, AVIRIS-NG was integrated into an Indian Space Research Organisation (ISRO) B-200 King Air for flights over India. 
Fifty-seven sites in India were selected for the Phase 1 campaign, which took place from December 2015 through March 2016. 
More than 300 flightlines were flown, covering a variety of forests, coastal margins, urban areas, agricultural areas, and sites of interest for geological and hydrological investigations. 
Approximate spatial resolution ranged between 3.3 m and 8.2 m, depending on the altitude of the aircraft and topography. 
Data were transformed from digital numbers to radiance units via standard practice calibration methods and corrections for spectral response function tails as in \cite{Thompson2018}. 

From the India dataset, we choose one flightline to highlight for its diverse set of test environments:
flightline \texttt{ang20160211t075004}, flown over Ahmedabad, includes a variety of land cover types and suspected methane sources. 
This flightline contains agricultural and urban land uses and has several low albedo water features.
The urban areas contain surfaces that correspond with spectral ``confusers'', as highlighted by~\cite{Ayasse2018}. 
Suspected methane sources include petroleum infrastructure, a landfill, and wastewater treatment plant tanks.

This flightline was processed with the three filtering methods: 
(1) the robust matched filter as a reference (``RMF''; sec.~\ref{sec:classicmf}), 
(2) the reweighted-$\ell_1$ matched filter method (``RWL1''; sec.~\ref{sec:sparse}), and
(3) the albedo-corrected reweighted-$\ell_1$ matched filter (``Albedo-Corrected RWL1''; sec.~\ref{sec:albedo}). 
These methods were applied on images partitioned such that the pixels from five adjacent detectors are processed in one filter instance.
Thirty iterations were used for the iterative methods.
As the total number of detectors is indivisible by five, the pixels from the fewer remaining detectors were processed together.
The target spectrum used for detection was generated following the description in section~\ref{sec:target}.

\subsection{Simulated Gas Enhancements}
\label{sec:simulation}

\begin{figure}[t!]
	\centering
	\includegraphics[width=\columnwidth]{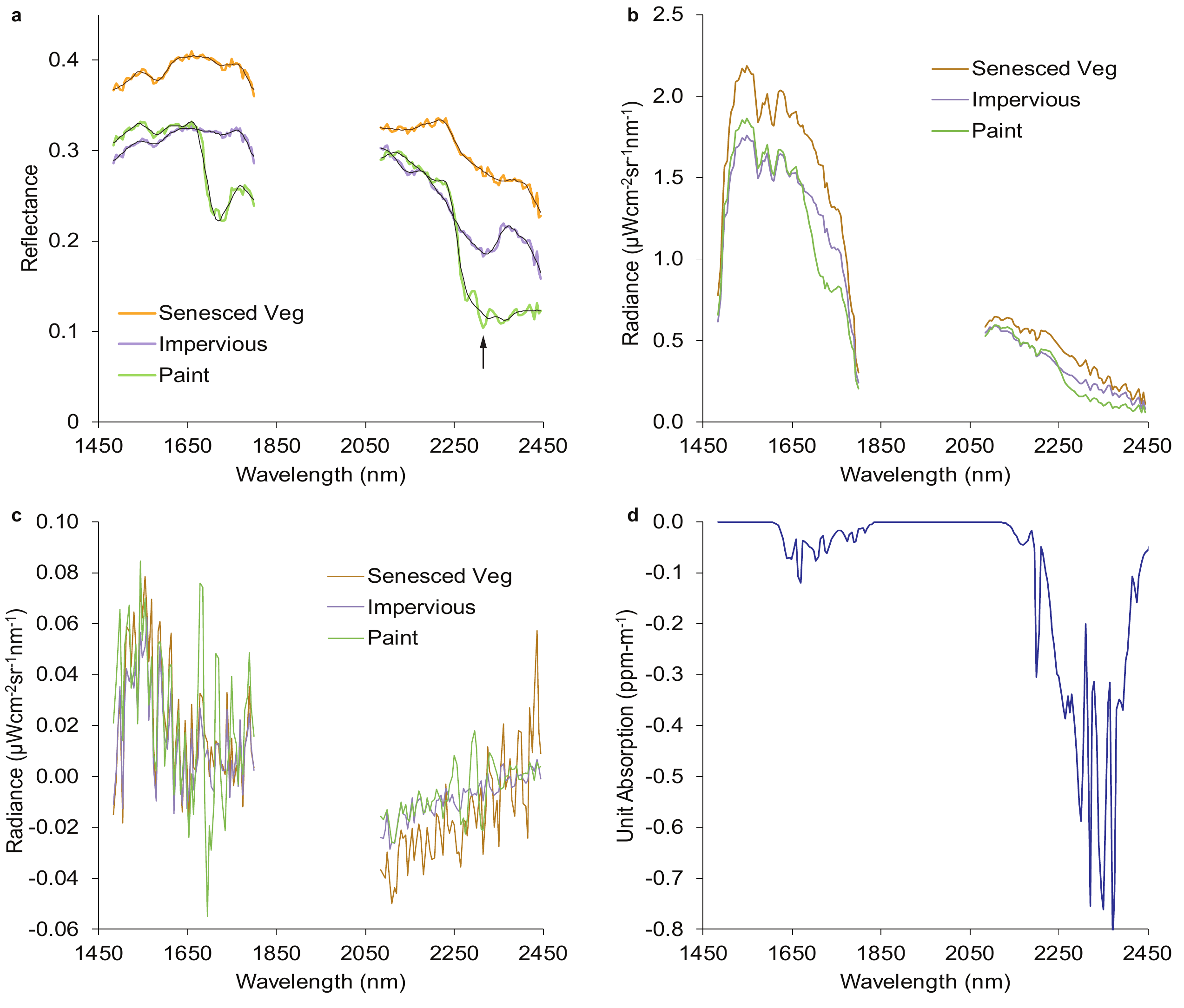}\vspace{-0.2cm}
	\caption{a) Reflectance for three example pixels from flightline \texttt{ang20160214t051014}. 
		Lighter background colors indicate original reflectance values, and darker colors indicate reflectance smoothed by Savitzsky-Golay filtering. 
		The arrow indicates a hydrocarbon absorption signature that was mostly removed by filtering. 
		b) Simulated radiance values for the same three pixels. 
		c) Residual between measured radiance and simulated radiance for the same three pixels. 
		The simulated radiance values have had artificial noise added.
		d) The unit absorption spectrum used for enhancement retrieval, representing the change in the natural log of radiance for a 1 ppm$\cdot$m change in methane concentration-pathlength.}
	\label{fig:simspectra}\vspace{-0.6cm}
\end{figure}

An AVIRIS-NG surface-reflectance image from an adjacent flightline \texttt{ang2016\-0214t051014} with no apparent methane plumes was used to create a validation radiance image with known methane concentration-pathlength values \cite{Dennison2013a,Ayasse2018}. 
Data from the flightline were atmospherically corrected to apparent surface reflectance by the Jet Propulsion Laboratory using the process described by \cite{Thompson2015a}. 
SWIR reflectance values between 1480 and 1800 nm and between 2080 and 2450 nm were extracted for smoothing; these wavelength ranges were selected to include methane absorption features but avoid water vapor and carbon dioxide absorption features. 
Reflectance values were spectrally smoothed using a fourth-degree polynomial Savitzky-Golay filter with a width of 21 bands to remove high-frequency signal and noise \cite{Savitzky1964,Schlapfer2011}. 
This smoothing process preserved broad spectral features like lignocellulose absorption \cite{Nagler2000}, but it eliminated narrow spectral features caused by atmospheric correction residuals and absorptions by surface materials that are easily confused with methane absorption \cite{Ayasse2018}. 
\figurename~\ref{fig:simspectra}a demonstrates how smoothing preserved broad absorption features expressed by surface materials. 
The three reflectance spectra shown are from example pixels containing senesced vegetation, which exhibits lignocellulose absorption; an urban impervious surface, which exhibits carbonate absorption; and a painted tennis court, which exhibits hydrocarbon absorption. 
Broader lignocellulose and carbonate absorption features were preserved by filtering, but a finer hydrocarbon absorption expressed by the painted surface, indicated by the arrow, was mostly eliminated by filtering (\figurename~\ref{fig:simspectra}a). 

MODTRAN6 simulations varying both surface albedo and methane con\-cen\-tra\-tion-pathlength were used to generate a reflectance-to-radiance lookup table \cite{Dennison2013a}, with MODTRAN inputs matching flightline parameters. 
An urban atmospheric profile with a visibility of 12 km was empirically found to most closely replicate measured radiance. 
For each band within each pixel, bilinear interpolation was used to determine the simulated radiance value from the lookup table, based on the reflectance value of that band and the concentration-pathlength enhancement, if present. 
Gaussian random noise was added to the radiance spectra \cite{Dennison2013a} based on a model of the AVIRIS-NG instrument, which includes photon and read noise for each band (R. Green, personal communication). 

Simulated radiance had a mean absolute error of 3.0\% at 2134 nm when compared to measured radiance (\figurename~\ref{fig:simspectra}b,c). 
Simulated radiance was mostly higher than the measured radiance from 1480 to 1800 nm and mostly lower from 2080 to 2450 nm (\figurename~\ref{fig:simspectra}c). 
Beyond those differences due to sensor noise, we attribute differences between simulated and measured radiance to aerosol scattering. 
MODTRAN urban aerosol properties may poorly approximate aerosols in a tropical urban environment. 
The linear shape of these residuals from 2080 to 2450 nm should result in negligible impacts on concentration-pathlength enhancement retrieval, due to the relatively fine spectral features of the methane unit absorption spectrum (\figurename~\ref{fig:simspectra}). 
 
Two simulated images were produced using flightline \texttt{ang20160214t051014}, using two different methods for determining concentration-pathlength enhancement. 
In the first simulated image, large eddy simulation (LES) was used to simulate realistic concentration-pathlength values based on an emission rate of 300 kg hr$^{-1}$ and a 4 m s$^{-1}$ surface wind \cite{Jongaramrungruang2018}. 
LES simulations were run at 5 m spatial resolution, vertically integrated to produce concentration-pathlength, and then spatially resampled to match the 8.1 m spatial resolution of this flightline. 
In the second simulated image, methane concentration-pathlength was enhanced in a randomly selected 1\% of pixels, with a uniformly random enhancement between 0 and 10000 ppm$\cdot$m. 
Enhancements in this image spanned a larger concentration-pathlength range than provided by the LES plumes.
The resulting simulated radiance images with known methane concentration-pathlength were processed to retrieve methane enhancements with the three filters using the same experimental procedure as for the unmodified Ahmedabad flightline,  but with no detector grouping.
For the randomly simulated enhancements, we also processed the image with additional matched filter variants with successive inclusion of iterative covariance estimation, albedo correction, and sparsity. 
The same simulated radiance image is used for all the filter algorithms within each simulation experiment (LES plume and random enhancement).

\section{Results}
\label{sec:results}

\subsection{Implementation Details}

We implemented each matched filter method in MATLAB. 
The albedo-correct\-ed reweighted $\ell_1$ method was also implemented in Python\footnote{Source code available: \url{https://www.github.com/markusfoote/mag1c}.}, using PyTorch for strong GPU acceleration \cite{NIPS2019_9015}. 
All processing was performed on a dual-socket Xeon E5-2640 2.60GHz workstation with 256 GB of memory and a Nvidia Quadro GV100 GPU.
Portions of each flightline censored by the Indian military were skipped during processing.

\subsection{Retrieval of Known Methane Point Sources}
\label{sec:resultsindia}
\begin{figure}[t!]  %
	\centering
	\includegraphics[width=\columnwidth,trim={0.9cm 0.3cm 0 0.25cm},clip]{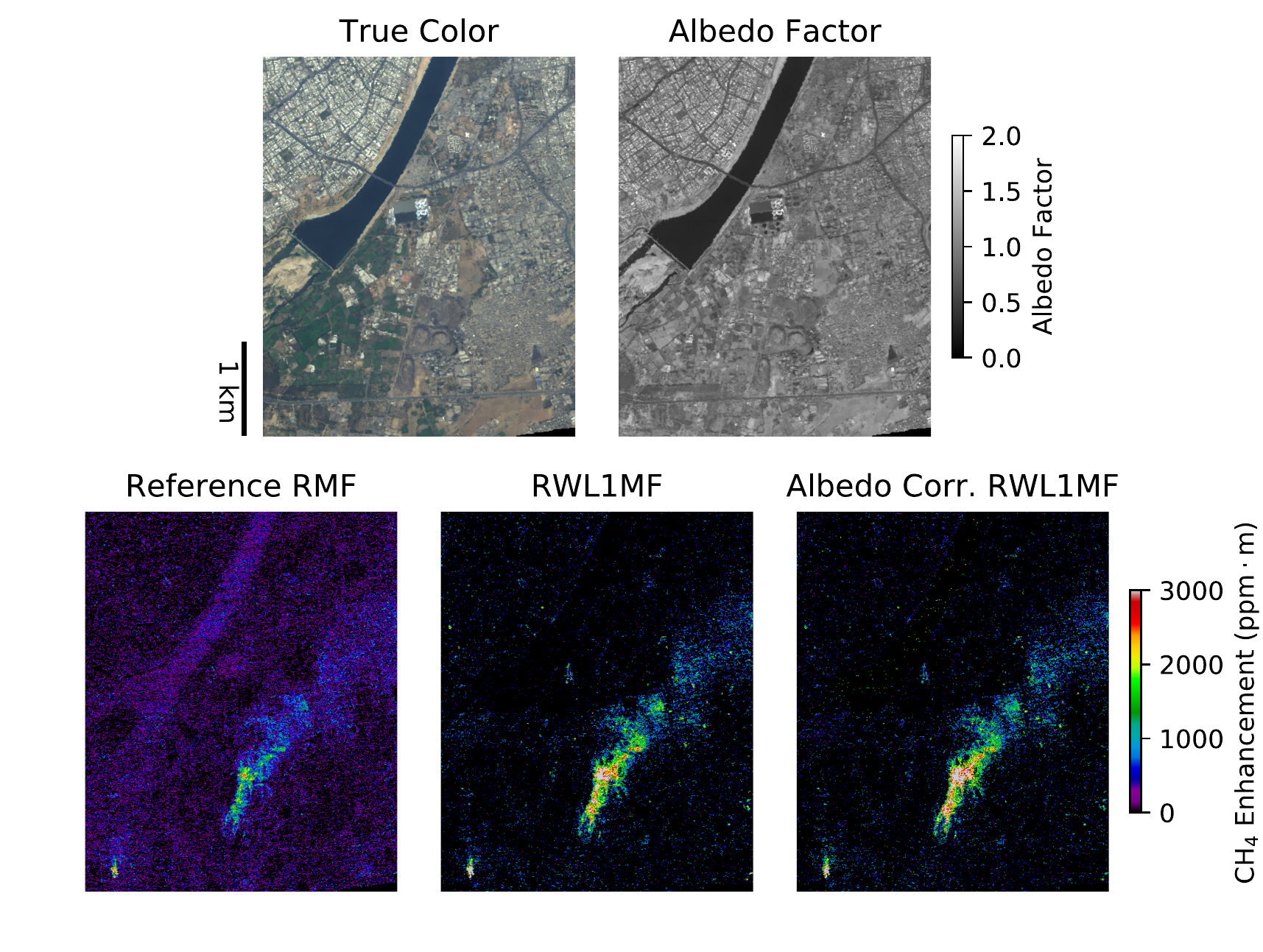}\vspace{-0.4cm}
	\caption{Matched filter results from a subset of flightline \texttt{ang20160211t075004}.
		The top row compares a true color composite from the radiance image and the albedo factor from the albedo-corrected RWL1 matched filter method.
		The bottom row compares the three matched filter methods. 
		Reference RMF: Reference Robust Matched Filter, RWL1MF: Reweighted-$\ell_1$ Matched Filter, Albedo-Corr. RWL1MF: Albedo-Corrected Reweighted-$\ell_1$ Matched Filter. }
	\label{fig:plumesgrid}
\end{figure}

The methane retrievals with both $\ell_1$ methods decreased background noise in the Ahmedabad flightline \texttt{ang20160211t075004} (\figurename~\ref{fig:plumesgrid}). 
Enhancements were more visible regardless of size due to increased contrast against the background. 
The standard deviation of a background region without obvious methane enhancements over a representative urban area was 
331.75~ppm$\cdot$m for the reference robust matched filter (``Reference RMF''), 
182.06~ppm$\cdot$m for the reweighted-$\ell_1$ matched filter (``RWL1''), and 
158.09~ppm$\cdot$m for the albedo-corrected reweighted-$\ell_1$ match\-ed filter (``Albedo-Corrected RWL1'').
The standard deviation for a background region with agricultural land cover was 
294.08~ppm$\cdot$m for the Reference RMF, 
129.61~ppm$\cdot$m for the RWL1MF, and
129.43~ppm$\cdot$m for the Albedo-Corrected RWL1MF.
The background noise level over water was notably lower for the RWL1 matched filter with a standard deviation of 
162.32~ppm$\cdot$m, compared to 
245.08~ppm$\cdot$m for the Albedo-Corrected RWL1MF and 
288.69~ppm$\cdot$m for the Reference RMF.
False positives remained similar across the matched filter results and coincided with surface features consistent with known `confuser' surfaces \cite{Ayasse2018}.

\begin{figure}[b!]  
	\centering
	\includegraphics[width=0.95\columnwidth,trim={0.1cm 0.25cm 0.55cm 0.25cm},clip]{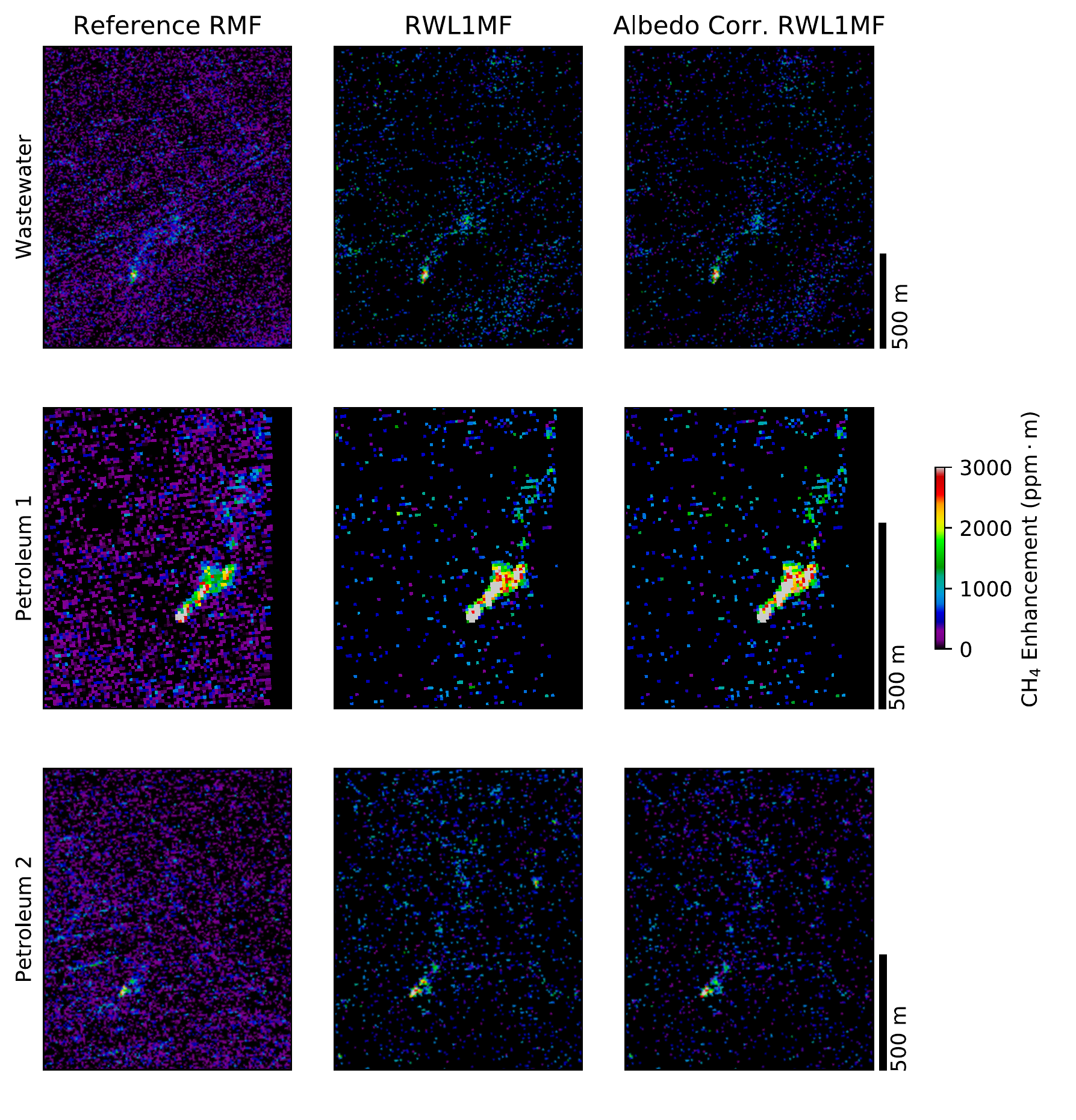}\vspace{-0.4cm}
	\caption{Example retrievals of methane plumes in \texttt{ang20160211t075004} using each retrieval method.}
	\label{fig:threeplumes}
\end{figure}

Five known plumes were contained within this flightline. 
We report findings about these plumes for the albedo-corrected reweighted $\ell_1$ matched filter.
The largest plume in \figurename~\ref{fig:plumesgrid} was from the Pirana landfill, which extended more than 5~km over urban areas and had a peak methane enhancement of 6750~ppm$\cdot$m. 
The smaller plume visible near the lower left corner of \figurename~\ref{fig:plumesgrid} was from a storage tank and extended about 700~m downwind of the source over agricultural land. 
The peak methane enhancement for this plume was 8835~ppm$\cdot$m. 
\figurename~\ref{fig:threeplumes} details the three remaining plumes. 
One plume was from a storage tank at a wastewater treatment plant and extended 570~m over the industrial areas of the treatment plant itself. 
The peak concentration within this plume was 6391~ppm$\cdot$m. 
The remaining two plumes were both from petroleum infrastructure. 
One plume extended 350~m over agricultural land before exiting the flightline's view with a peak enhancement of 24681~ppm$\cdot$m.
The other plume extended 250~m over an urban area with 7255~ppm$\cdot$m peak enhancement.

\begin{figure}[t!]
	\centering
	\includegraphics[width=\columnwidth]{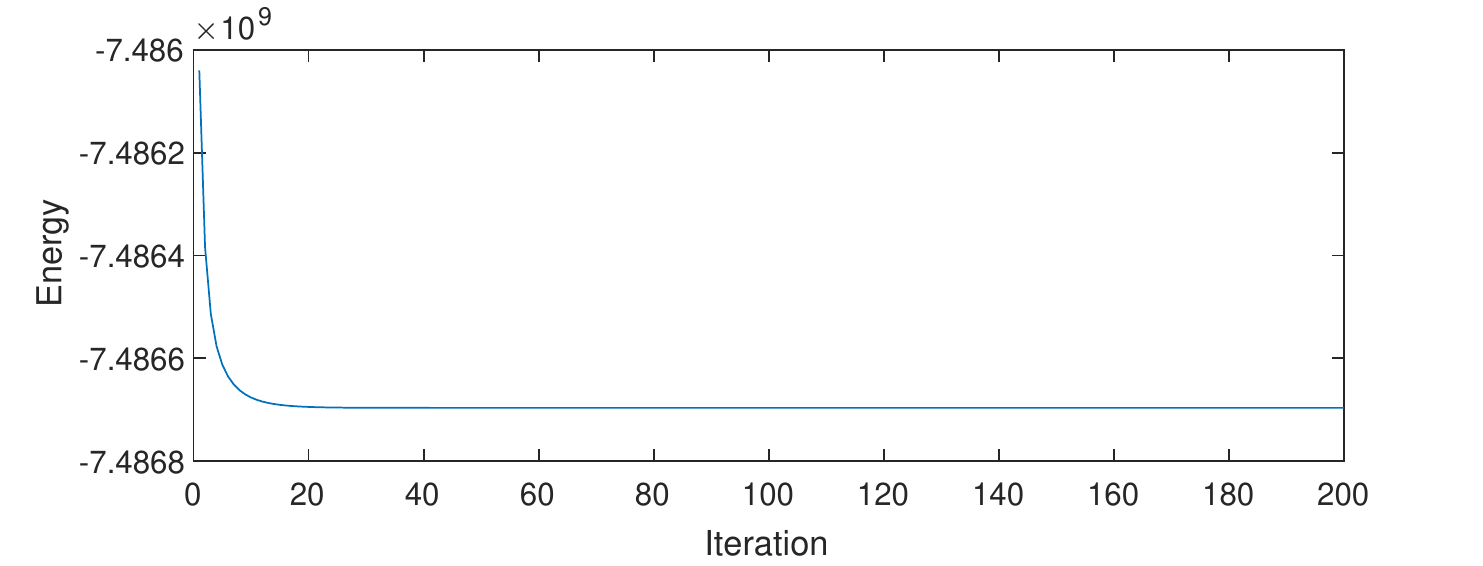}\vspace{-0.3cm}
	\caption{Energy of Albedo-Corrected Reweighted-$\ell_1$ optimization for the Ahmedabad flightline when run with all detectors grouped and 200 iterations.}
	\label{fig:energy}\vspace{-0.4cm}
\end{figure}

We also show the convergence properties of the albedo-corrected reweighted-$\ell_1$ matched filter algorithm using this flightline.
The energy defined by (\ref{eq:spatialrwl1}) was tracked during 200 iterations of the algorithm, with all 598 detectors processed together (\figurename~\ref{fig:energy}).
The optimization attained stable convergence within 20 iterations. 
All other experiments are run for 30 iterations, providing an additional margin on convergence.

\subsection{Simulated Plume Retrievals}
\label{sec:simplume}
\begin{figure}[tbp]  
	\centering
	\includegraphics[width=0.95\columnwidth,trim={1.5cm 0.1cm 0.9cm 0.1cm},clip]{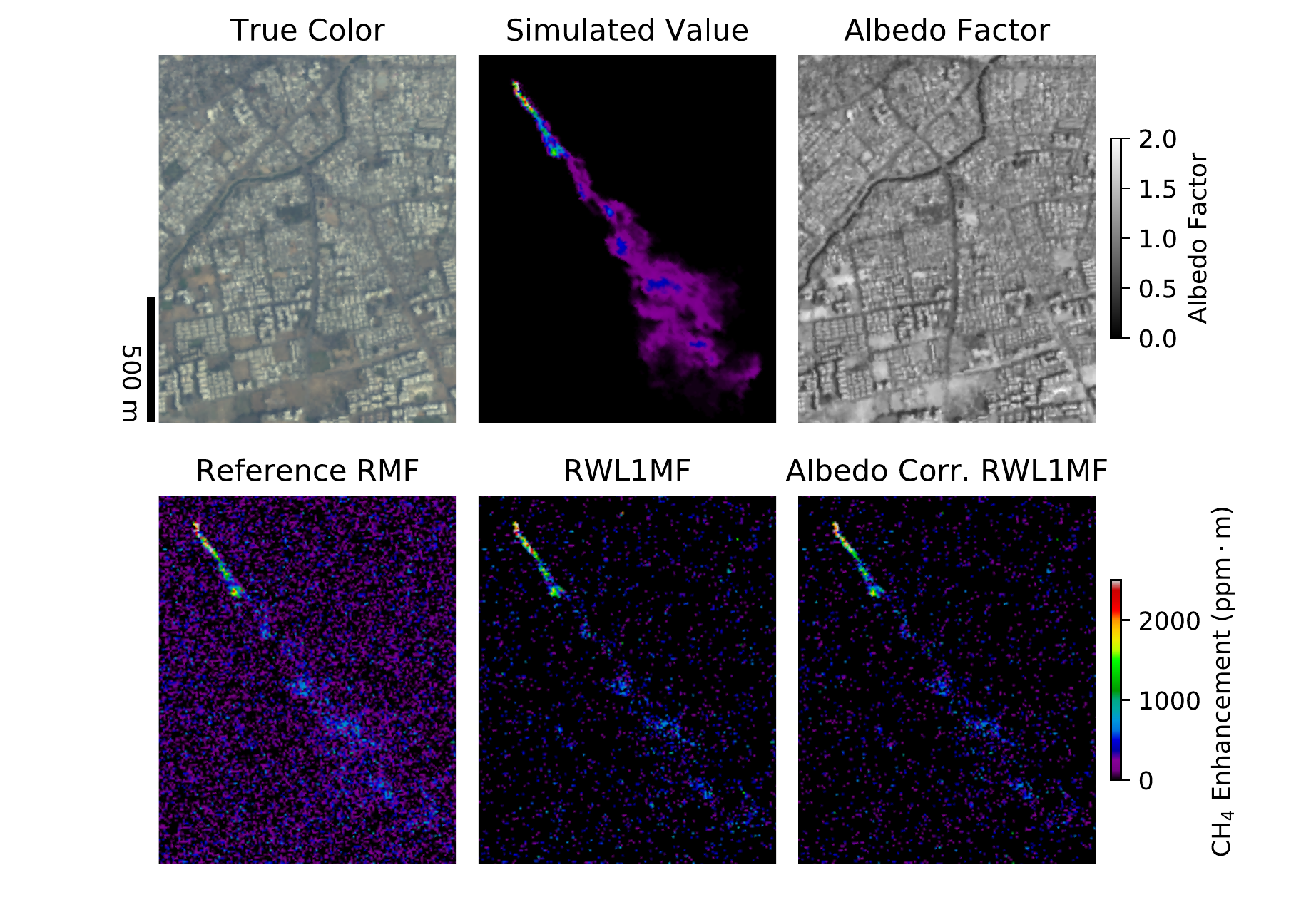}\vspace{-0.5cm}
	\caption{Example retrievals of simulated methane plumes using each retrieval method.}
	\label{fig:simview}
\end{figure}

\begin{figure}[tbp]  
	\centering
	\includegraphics[width=1.0\columnwidth]{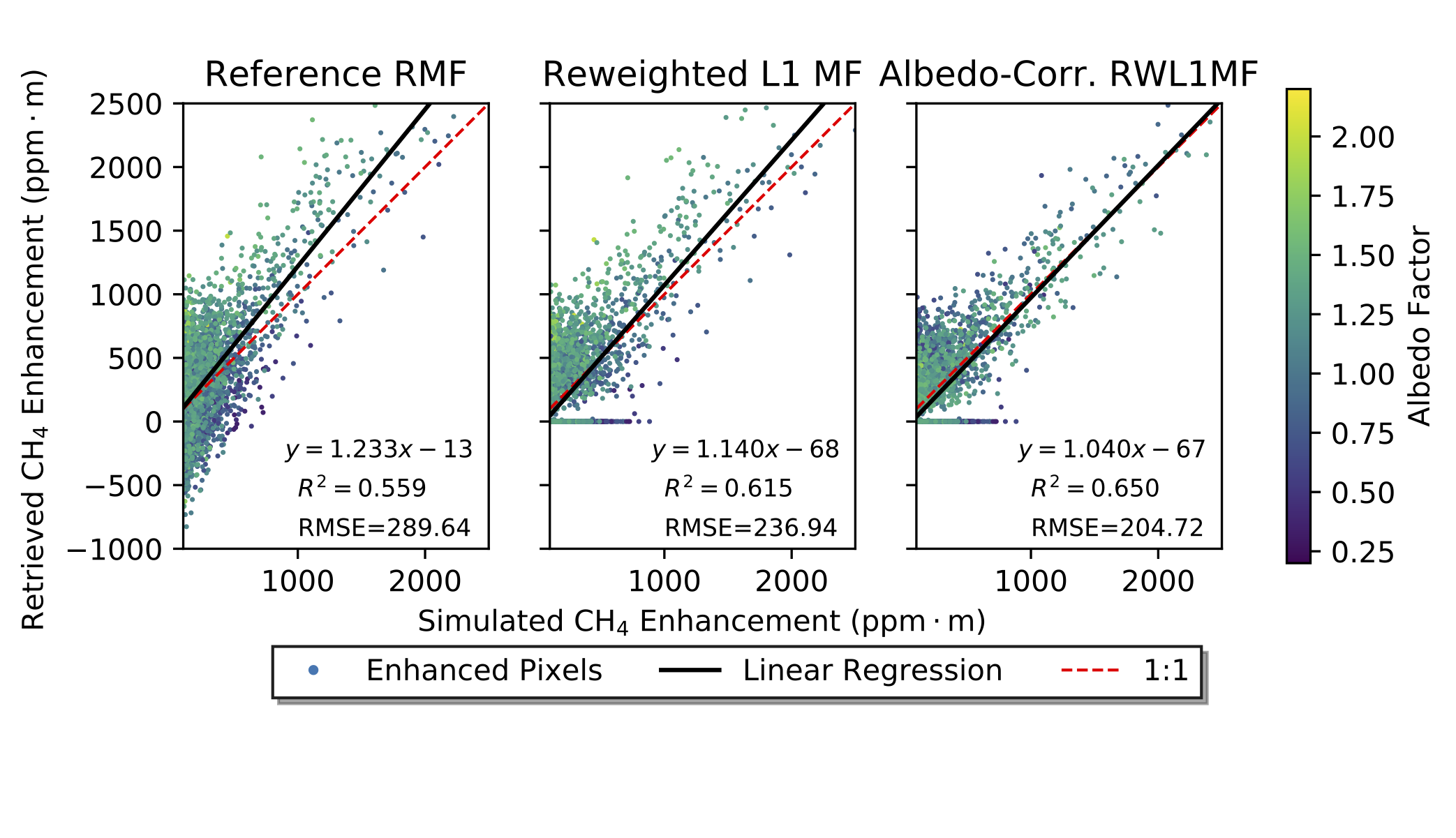}\vspace{-0.7cm}
	\caption{Retrieval of simulated methane enhancements for the plume overlay simulated image. 
	Each point represents a single pixel with simulated and retrieved enhancement value.
	A linear regression for each algorithm is shown in black.
	Only pixels with a simulated enhancement above the approximate background noise of 100 ppm$\cdot$m are included in these plots and regressions. Red dashed lines have slope 1 and no offset for reference.}
	\label{fig:simplots}
\end{figure}

\begin{figure}[tbp] 
	\centering
	\includegraphics[width=1.0\columnwidth]{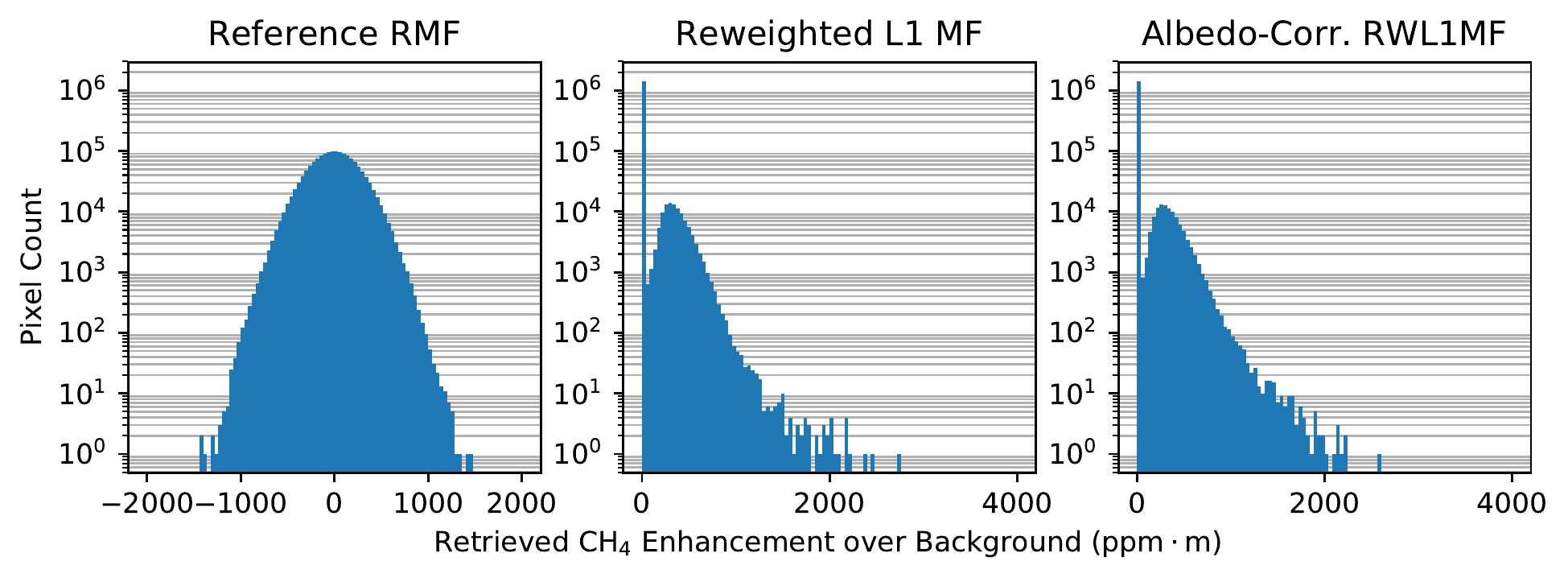}\vspace{-0.25cm}
	\caption{Histograms for the nonenhanced pixels from the plume overlay simulation for each retrieval method show the distribution of background retrieved enhancement.
	Note the logarithmic scale of the pixel counts -- the Reference RMF distribution is Gaussian.}
	\label{fig:simhist}
\end{figure}

Simulated plumes overlaid on an image otherwise free of obvious methane enhancement (\texttt{ang20160214t051014}, also over Ahmedabad) provided a realistic retrieval situation with known true concentration-pathlength.
Overlaid simulated plumes appeared more clearly with both $\ell_1$ methods than with the reference robust matched filter (\figurename~\ref{fig:simview}).
The retrieved enhancement concentration-pathlength for each pixel was compared to the simulation value in \figurename~\ref{fig:simplots}. 
This comparison revealed the ability of sparse matched filter methods to significantly reduce background noise.
The standard deviation of background pixels with no simulated enhancement over agricultural land was 
238.8~ppm$\cdot$m for the Reference RMF, 
86.58~ppm$\cdot$m for the RWL1MF, and %
90.35~ppm$\cdot$m for the Albedo-Corrected RWL1MF. %
These values represent a reduction to 36.3\% and 37.8\% of the Reference RMF agricultural background standard deviation for RWL1MF and Albedo-Corrected RWL1MF agricultural background standard deviations, respectively.
This background standard deviation for Albedo-Corrected RWL1MF is a factor of 2.64 reduction over the Reference RMF agricultural background standard deviation.
The standard deviation of background pixels with no simulated enhancement over urban areas was
255.1~ppm$\cdot$m for the Reference RMF, 
106.7~ppm$\cdot$m for the RWL1MF, and %
105.5~ppm$\cdot$m for the Albedo-Corrected RWL1MF. %
These values represent a reduction to 41.8\% and 41.3\% of the Reference RMF urban background standard deviation for RWL1MF and Albedo-Corrected RWL1MF background standard deviations, respectively.
The background standard deviations for this simulated plume image are 58\% (urban RWL1) to 81\% (agricultural Reference RMF) of the values from the previous real flightline in section~\ref{sec:resultsindia}.
A small bias of the reweighted $\ell_1$ methods toward lower concentration retrievals was evident in the lower slopes for the linear regressions on pixels with simulated enhancements.
Pixels with extreme albedo factors tended to have higher error, although this error was reduced by the albedo correction.
Enhancements were recovered most reliably with the albedo-corrected filter with the highest coefficient of determination.
The histograms of simulated nonenhanced background pixels showed that both reweighted-$\ell_1$ methods produced lower background enhancements over a smaller number of pixels relative to the Reference RMF (\figurename~\ref{fig:simhist}).

\subsection{Random Simulated Enhancement Retrievals}
\label{sec:random}

The randomly simulated enhancement revealed algorithm behavior over a larg\-er enhancement range -- up to 10,000~ppm$\cdot$m -- and with uniform sampling of methane enhancement. 
The proposed algorithm incorporates three separate improvements over the reference RMF\cite{Thompson2015}: albedo correction, iterative with positivity constraint, and sparse $\ell_1$ prior. We now present results showing the impact of these individual improvements.

This random simulation image was processed with three additional filter variants along with the three previous filter methods.
First, an albedo-only (traditional) matched filter demonstrates the effects of albedo correction alone.
The second and third additional filters are iterative with positivity constraint without a sparsity term.
This iterative filter is used both with and without albedo correction.
These iterative methods demonstrate the effects of the iterative estimation with a positivity constraint independent of sparsity (and albedo correction, for the variant without albedo correction).
The features included in each algorithm are summarized in Table \ref{tab:algs}.
The application of an albedo correction factor for all algorithms is controlled by restricting all $r_i = 1$ in the algorithm provided in Fig.~\ref{alg:spatialrwl1mf}. 
The iterative with positive constraint algorithms exclude the sparse prior by setting $\boldsymbol{w} = 0$ in Fig.~\ref{alg:spatialrwl1mf}. 
The positivity constraint of the iterative algorithms (both with and without sparsity) is applied within each iteration, before the background mean and covariance are calculated. 
The positivity constraint is necessary in the iterative algorithms to prevent adding a methane signature to the covariance matrix and instead only remove any methane signature that exists within the original radiance data.

\begin{table}[b]
	\renewcommand{\arraystretch}{1.1}
	\setlength{\tabcolsep}{4.5pt}
\newcommand*{\boldcheckmark}{%
  \textpdfrender{
    TextRenderingMode=FillStroke,
    LineWidth=.3pt, 
  }{\checked}%
}
	\newcommand{\cmark}{\boldcheckmark}%
	\newcommand{\xmark}{$\boldsymbol{\times}$}%
	\caption{Feature Summary for Retrieval Algorithms in Section \ref{sec:random}}
	\label{tab:algs}
	\centering%
	\begin{tabular}{ccccc}
	\toprule
	 & Iterative with & Sparse & Albedo \\
	 Algorithm & Positive Constraint & Prior & Correction \\
	\midrule
	Reference RMF \textsuperscript{\textdagger}		   & \xmark & \xmark & \xmark \\
	Albedo-Only MF 									   & \xmark & \xmark & \cmark \\
	Iter. Pos. w/o Albedo MF						   & \cmark & \xmark & \xmark \\
	Iter. Pos. w/ Albedo MF						       & \cmark & \xmark & \cmark \\
	RWL1MF \textsuperscript{\textdagger}			   & \cmark & \cmark & \xmark \\
	Albedo-Corr. RWL1MF \textsuperscript{\textdagger}  & \cmark & \cmark & \cmark \\
	\bottomrule
	\end{tabular}\\\vspace{1pt}%
	\cmark~--~Feature included in algorithm. \xmark~--~Unused feature.\\
	\textsuperscript{\textdagger}Algorithm also used in Section \ref{sec:simplume}.
\end{table}

\begin{figure}[p!]  %
	\centering
	\includegraphics[width=\columnwidth,clip,trim={0 1.7cm 0 0}]{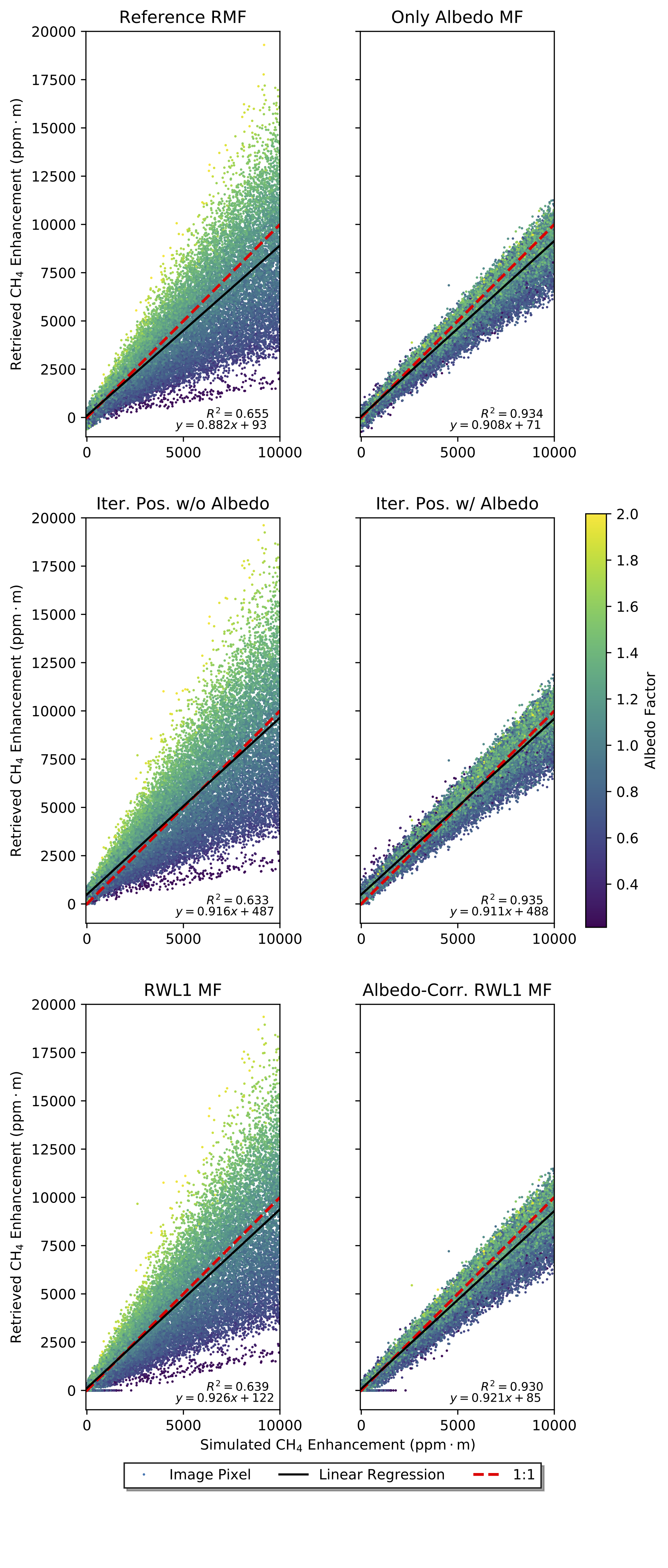}
	\vspace{-0.7cm}\caption{Scatterplots comparing simulated and retrieved methane enhancements for randomly placed concentration-pathlength values ranging from 0 to 10000 ppm$\cdot$m. 
	Only pixels with a nonzero simulated enhancement are included in these plots.
	Inset plots magnify the portion of each plot from 0 to 2500 ppm$\cdot$m.  
	Points across all plots are colored according to the albedo factor value from the albedo corrected methods.
	Black lines are linear regressions of the points in each plot. 
	Red dashed lines have slope 1 and no offset for reference.
	}
	\label{fig:random}
\end{figure}
Retrieved methane enhancement values for pixels that had a simulated enhancement above background are shown in Fig.~\ref{fig:random}.
Linear regressions of these enhanced pixels are shown on the scatterplot for the result of each algorithm.
Each horizontal pair within Fig.~\ref{fig:random} shows the effect of adding albedo correction to each corresponding algorithm.
Albedo correction reduced the variance of retrieved methane concentration for retrievals of simulated enhancement.
The effects of albedo correction, and the problem presented by varied surface albedo, are apparent in the vertical color gradient of points in the non-albedo-corrected plots that is less apparent in the plots produced from albedo-corrected algorithms.
The trend of low albedo causing low retrieval (and high albedo causing high retrieval) was reduced in the methods with albedo correction.
Even pixels with a low albedo factor (\textasciitilde0.2\,--\,0.3 of mean radiance) that are notable outliers in the Reference RMF and RWL1MF results are corrected and appeared within the main distribution of retrievals in the albedo-corrected methods. 
Retrievals between 0-150~ppm$\cdot$m are infrequent in the reweighted-$\ell_1$ methods from the sparsity prior and a collection of zero-valued retrievals appear along the x-axis in these algorithms. 
This gap is reduced with albedo correction and is not present in the Reference RMF or Albedo-Only MF. 
The iterative positive-constrained algorithms have no negative retrieval values, but the bias of retrievals is increased. 
The albedo-corrected (traditional) matched filter has the lowest bias (71 ppm$\cdot$m), while the albedo-corrected reweighted-$\ell_1$ algorithm (85 ppm$\cdot$m) and Reference RMF (93 ppm$\cdot$m) have similar biases.
Although the iterative with positive constraint algorithm with albedo correction produced the best $R^2$ for simulated-enhancement pixels, this algorithm also had the largest bias (488 ppm$\cdot$m). Other albedo-corrected algorithms had very similar $R^2$ values.
The reweighted-$\ell_1$ filter with albedo correction produced the second most accurate values for both regression bias and slope, with $R^2$ comparable to the other well-performing algorithms.
 
\begin{figure}[tbp]
		\centering
		\includegraphics[width=0.9\columnwidth]{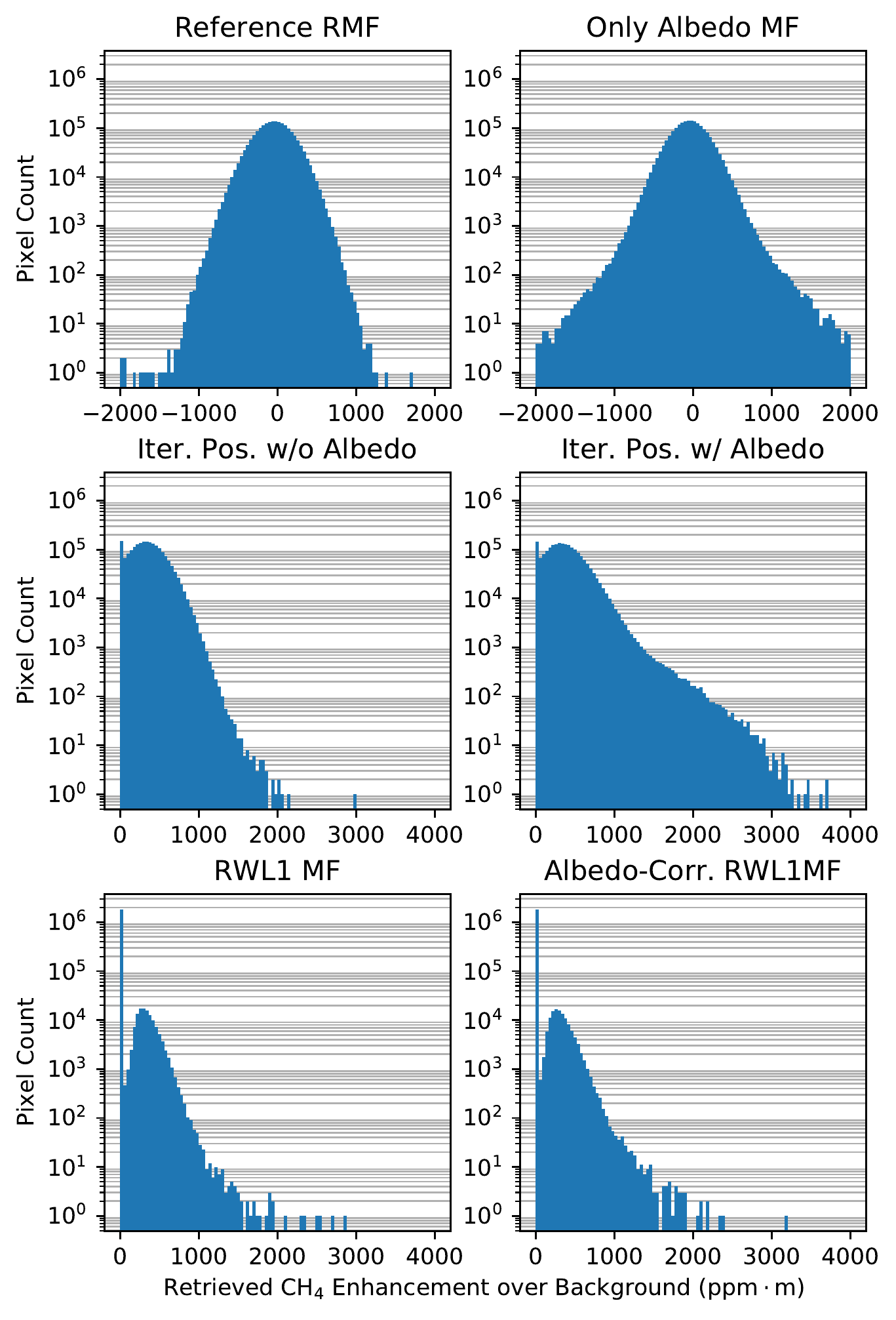}
		\caption{Histograms for the nonenhanced pixels from the random enhancement simulation for each retrieval method show the distribution of background retrieved enhancement.
		Note the logarithmic scale of the pixel counts.}
		\label{fig:randomhist}
\end{figure}
The behavior of methane concentration retrievals for pixels with no simulated methane enhancement is summarized in Fig.~\ref{fig:randomhist}. 
The iterative with positivity constraint algorithms had pixel counts of zero enhancement similar to the traditional matched filter methods. 
Reweighted-$\ell_1$ sparsity showed a much larger increase in the number of pixels with retrieved zero enhancement over the iterative with positivity constraint algorithms and an order of magnitude decrease in the number of pixels with low retrieved enhancement.
Albedo correction generally widened the distribution of methane enhancement retrievals for nonenhanced pixels.
Of the pixels with zero simulated enhancement, the Albedo-Corrected RWL1MF algorithm exactly estimated zero enhancement for 93.9\%. 

We quantified the error in retrieved methane enhancement from the simulated value by the root mean squared error (RMSE). 
Three RMSE values were calculated for each algorithm's result in Table \ref{tab:rmse}. 
The errors of the enhanced and non-enhanced simulated values were calculated separately. 
RMSE for pixels with a methane enhancement (pixels shown in Fig.~\ref{fig:random}) is shown in column two. 
RMSE for pixels without a methane enhancement (pixels represented in Fig.~\ref{fig:randomhist}) is shown in column three. 
Additionally, the total RMSE for all pixels, both those with and without simulated methane enhancement, is shown in column four, with the relative improvement of the all-pixel RMSE over the Reference RMF method in column five. 
For the pixels with a simulated methane enhancement, the RMSE of all albedo-corrected algorithms decreased from the corresponding algorithm without albedo correction. 
The RMSE remained similar or increased for the nonenhanced pixels with the inclusion of albedo correction. 
When using iterative statistic estimations with a positivity constraint, the errors of enhanced pixels remained similar while nonenhanced pixel errors nearly doubled.
Although RMSE of enhanced pixels slightly increased, the RMSE for nonenhanced pixels significantly decreased for the albedo-corrected reweighted-$\ell_1$ algorithm. 
The all-pixel RMSE captured the overall performance of each algorithm without classification of enhanced pixels. 
The best RMSE was achieved with the albedo-corrected sparse matched filter with a relative improvement over the Reference RMF of 60.7\%.
\begin{table}[bp]
	\renewcommand{\arraystretch}{1.3}
	\setlength{\tabcolsep}{1.5pt}\vspace{-0.2cm}
	\caption{RMSE Values of Retrieval Algorithms}
	\label{tab:rmse}
	\centering%
	\begin{tabular}{cS[table-format=4.3]S[table-format=3.3]S[table-format=3.3]S[table-format=2.1,table-sign-mantissa=true,table-space-text-post = \%]}
	\toprule
	 &  \multicolumn{3}{c}{RMSE (ppm$\cdot$m)} \\
	\cmidrule(lr){2-4}
	 & {Enhanced} & {Non-Enhanced} & {All} & {Relative}\\
	{Algorithm} & {Pixels} & {Pixels} & {Pixels} &{Improvement}\\
	\midrule %
	Reference RMF 			& 1951.075 &  237.188 &  306.094 & 0.0\% \\ 
	Albedo-Only MF 			&  842.549 &  240.492 &  253.663 & 17.1\% \\
	Iter. Pos. w/o Albedo MF	& 2035.088 &  404.754 &  451.143 & -47.3\% \\
	Iter. Pos. w/ Albedo MF &  743.802 &  450.124 &  453.995 & -48.3\% \\
	RWL1MF 					& 2040.876 &   90.935 &  223.073 & 27.1\% \\
	Albedo-Corr. RWL1MF 	&  826.744 &   87.759 &  120.197 & 60.7\% \\
	\bottomrule
	\end{tabular}
\end{table}

\begin{figure}[t!] 
	\centering
	\includegraphics[width=0.95\columnwidth]{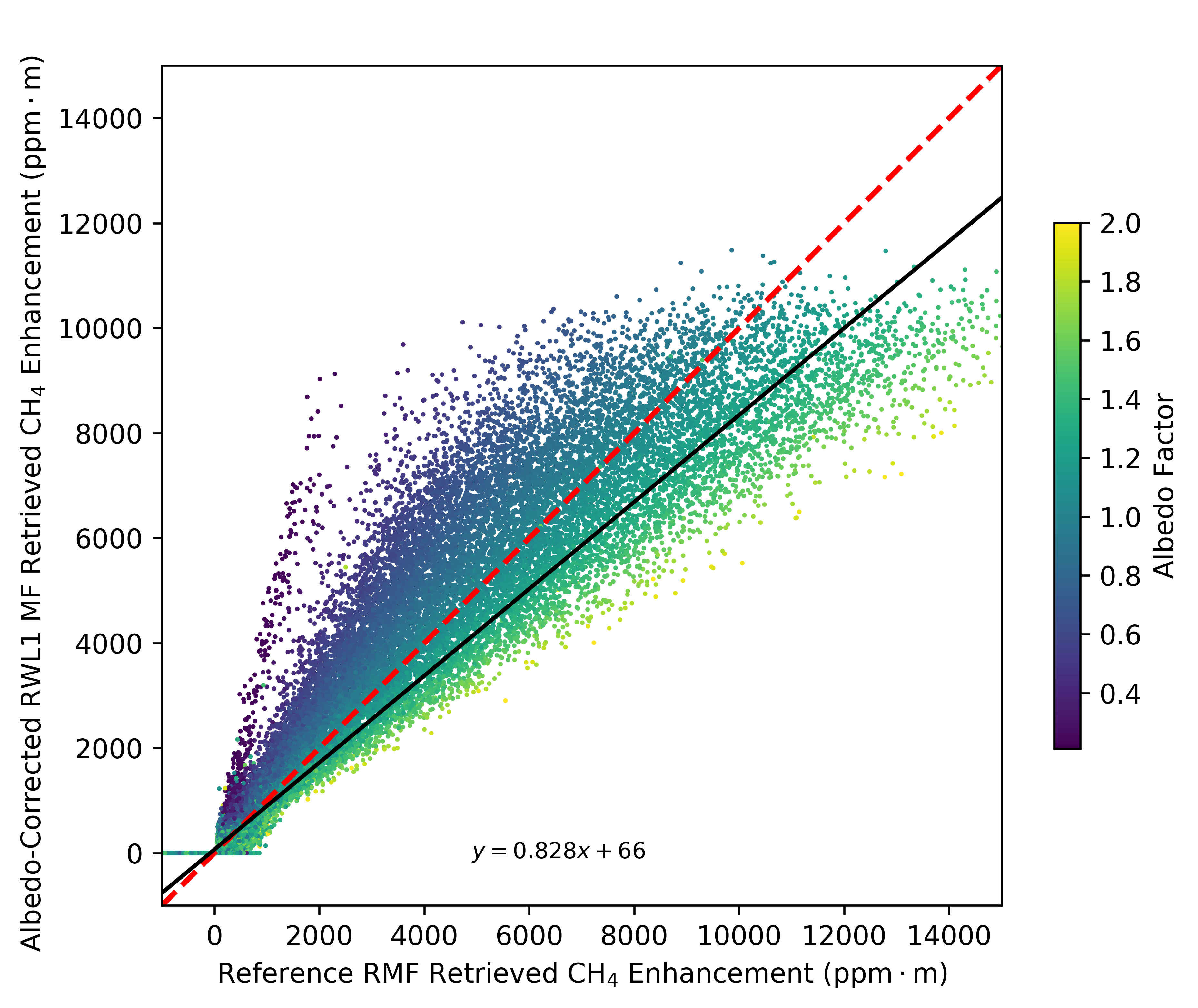}
	\caption{Scatterplot of Retrieved Methane Enhancement from Albedo-Corrected RWL1 MF method versus from Reference RMF methods. The dashed red line indicates a 1:1 relationship. The black line is a linear regression of the data, with the equation of this best-fit line shown on the plot.}
	\label{fig:versus}
\end{figure}

The cumulative effect of the iterative, positivity-constraint, sparsity, and albedo-correction contributions over the Reference RMF is summarized in Fig.~\ref{fig:versus}. The overall trend of these contributions is to estimate a lower enhancement than was produced by the reference matched filter method. 
The change in retrieved enhancement was dependent on the underlying albedo of the pixel. For pixels with a sufficiently low albedo, the retrieval was increased. Negative retrievals by the Reference RMF were mapped to zero in the albedo-corrected sparse result.

\subsection{India Dataset Processing}

We processed the entire India Phase 1 campaign dataset with the Albedo-Corrected Reweighted-$\ell_1$ MF algorithm. 
The complete dataset, 2.4~TB of radiance files for 300 flightlines, was processed in 8.3~hours using the GPU for computational acceleration in a desktop workstation. 
Solely reading (and not processing) the entire dataset under ideal conditions from hard disk storage at 100~MBps would theoretically take 6.6~hours. 
The total time to completion was limited by the speed of reading the dataset from storage, which indicates that the algorithm is suited for low-latency production of methane detection data products for both real-time applications and scalability to large flight campaigns and ground processing of data from future spaceborne imaging spectrometers.

\section{Discussion}
\label{sec:discussion}

Methane enhancement retrieval is improved over low albedo surfaces with the albedo-corrected reweighted $\ell_1$ matched filter algorithm. 
This improvement is implemented with minimal computational expense over the similar reweighted $\ell_1$ matched filter.
Surfaces with albedo factors as low as 0.2 (1/5{th} of the mean image partition radiance) are consistently corrected to retrieve methane enhancements more accurately.
The random enhancement analysis (Sec.~\ref{sec:random}) identifies sparsity and albedo correction as complementary advancements to increase the accuracy of trace gas enhancement retrievals. 
The cumulative effect of these contributions for RMSE reduction is greater than the sum of each contribution alone.
As separable contributions, applications that have computational constraints could use only albedo correction to capture the associated improvement on retrieval accuracy, without $\ell_1$ optimization or robust statistics calculation.
This would have the additional benefit of preserving Gaussianity in the retrieval noise statistics for the purposes of uncertainty assessment and propagation.

Images containing simulated plumes allowed a direct comparison between actual and retrieved enhancements, which is not typically possible using imaging spectrometer data acquired over real plumes. 
Despite including a radiance-dependent noise function provided by JPL, the background standard deviation from the matched filters applied to the simulated plume image was 58-81\% of that from matched filters applied to the adjacent real flightline.
This noise model may not account for all noise present within the real data. 
Despite lower error in the simulated image, our results still consistently show that the sparse prior and albedo correction reduced the background standard deviation to between 36.3\%-47.1\% that of the reference robust matched filter in both real and simulated data. 

Accurate retrievals are also limited by variables that vary within and between flightlines but are not accounted for in this work. 
The method we used for generating a unit absorption spectrum relies on radiative transfer simulations based on specific geometric and atmospheric parameters. 
The change in radiance with respect to a change in concentration-pathlength will vary with parameters such as solar zenith angle, ground elevation, and atmospheric water vapor concentration. 
For more accurate retrieval of the concentration-pathlength enhancement, the unit absorption spectrum must thus be "tuned" for individual images. 
For airborne remote sensing across regions spanning tens of kilometers, a uniform unit absorption spectrum may adequately represent variation in geometric and atmospheric parameters. 
For potential future satellite missions, use of a spatially varying unit absorption spectrum will be essential to enhancement retrieval. 
To account for spatially varying water vapor, water vapor abundance could be estimated using traditional single-spectrum techniques (e.g. \cite{Thompson2018}). 
Analysis of errors in enhancement retrieval resulting from unit absorption spectrum assumptions should be a high priority for future work. 
The ability of a uniform unit absorption spectrum to detect methane enhanced pixels should not be affected by errors in geometric and atmospheric parameters, even if these errors result in a systematic bias in retrieved concentration-pathlength. 

Both the target spectrum and sparse prior are agnostic to the specific gas species and wavelength range.
Our target spectrum generation method is applicable to any gas with distinct absorption features, including other greenhouse gasses, such as CO$_2$, that have been successfully mapped using AVIRIS and AVIRIS-NG \cite{Dennison2013a,Dennison2013b,Thorpe2017}.
Both our target spectrum generation and sparse prior optimization methods are directly applicable to trace gas retrievals using data from other hyperspectral instruments.
Our improvements impact other work in the remote sensing of methane, including flux estimation \cite{Frankenberg2016}, which depends on accurate concentration-pathlength values. 
Additionally, automated plume detection with machine learning will benefit from the decreased background noise.

\section{Conclusion}
\label{sec:conclusion}

Methane enhancement retrievals based on a matched filter method benefit from albedo correction and sparse prior.
The sparse prior enforces the expectation that methane enhancements are rare events within large flightlines. 
The albedo correction provides more accurate concentration estimates over surfaces of low reflectance.
This approach requires no hyperparameters to be selected beyond the number of iterations, and good convergence is achieved after 20 iterations.
The iterative method that is required for sparse optimization ensures that the target signal does not corrupt the covariance matrix \cite{Theiler2006a}.
These improvements produce methane enhancements that more accurately predict the true enhancement values in validation datasets. 
Additionally, methane plumes from point sources are more discernible against the background noise with the background standard deviation decreased by a factor of over 2.6 in the validation dataset.

\section*{Acknowledgment}
We acknowledge NASA Earth Science Division sponsorship of the AVIRIS-NG instrument and the efforts of the AVIRIS-NG team on the India Campaign. 
We are grateful for the support of NVIDIA Corporation by providing the GPU used for this research.
A portion of this research was performed at the Jet Propulsion Laboratory, California Institute of Technology, under contract with the National Aeronautics and Space Administration (NASA). %

\subsection*{Data Statement}
AVIRIS-NG data are available upon request from JPL. 
See \url{https://aviris-ng.jpl.nasa.gov}. 
Other data are available upon request from the authors.

\ifCLASSOPTIONcaptionsoff
\newpage
\phantom{A}
\newpage

\setstretch{2} 

\processdelayedfloats

\newpage 
\fi

\bibliographystyle{IEEEtran}
\bibliography{IEEEabrv,gasMap}

\begin{thebibliography}{10}
\providecommand{\url}[1]{#1}
\csname url@samestyle\endcsname
\providecommand{\newblock}{\relax}
\providecommand{\bibinfo}[2]{#2}
\providecommand{\BIBentrySTDinterwordspacing}{\spaceskip=0pt\relax}
\providecommand{\BIBentryALTinterwordstretchfactor}{4}
\providecommand{\BIBentryALTinterwordspacing}{\spaceskip=\fontdimen2\font plus
\BIBentryALTinterwordstretchfactor\fontdimen3\font minus
  \fontdimen4\font\relax}
\providecommand{\BIBforeignlanguage}[2]{{%
\expandafter\ifx\csname l@#1\endcsname\relax
\typeout{** WARNING: IEEEtran.bst: No hyphenation pattern has been}%
\typeout{** loaded for the language `#1'. Using the pattern for}%
\typeout{** the default language instead.}%
\else
\language=\csname l@#1\endcsname
\fi
#2}}
\providecommand{\BIBdecl}{\relax}
\BIBdecl

\bibitem{Saunois2016}
M.~Saunois \emph{et~al.}, ``{The global methane budget 2000-2012},''
  \emph{Earth Syst. Sci. Data}, vol.~8, no.~2, pp. 697--751, 2016. doi:
  10.5194/essd-8-697-2016.

\bibitem{Myhre2013}
G.~Myhre \emph{et~al.}, ``Anthropogenic and {Natural} {Radiative} {Forcing},''
  in \emph{Climate {Change} 2013 - {The} {Physical} {Science} {Basis}},
  {Intergovernmental Panel on Climate Change}, Ed.\hskip 1em plus 0.5em minus
  0.4em\relax Cambridge: Cambridge University Press, 2014, pp. 659--740, doi:
  10.1017/CBO9781107415324.018.

\bibitem{Kirschke2013}
S.~Kirschke \emph{et~al.}, ``{Three decades of global methane sources and
  sinks},'' \emph{Nature Geoscience}, vol.~6, no.~10, pp. 813--823, 2013. doi:
  10.1038/ngeo1955.

\bibitem{Ciais2013}
P.~Ciais \emph{et~al.}, ``{Carbon and Other Biogeochemical Cycles},'' in
  \emph{Climate Change 2013 - The Physical Science Basis}, 2014, pp. 465--570.

\bibitem{IPCC2016}
{United States Environmental Protection Agency},
  ``\BIBforeignlanguage{en}{Inventory of {U}.{S}. {Greenhouse} {Gas}
  {Emissions} and {Sinks}: 1990-2014},'' Reports and {Assessments} EPA
  430-R-16-002, Apr. 2016.

\bibitem{Nisbet2014}
E.~G. Nisbet, E.~J. Dlugokencky, and P.~Bousquet, ``{Methane on the
  Rise--Again},'' \emph{Science}, vol. 343, no. 6170, pp. 493--495, 2014. doi:
  10.1126/science.1247828.

\bibitem{Nisbet2019}
E.~G. Nisbet \emph{et~al.}, ``{Very Strong Atmospheric Methane Growth in the 4
  Years 2014-2017: Implications for the Paris Agreement},'' \emph{Global
  Biogeochem. Cycles}, vol.~33, no.~3, pp. 318--342, Mar. 2019. doi:
  10.1029/2018GB006009.

\bibitem{Nisbet2016}
------, ``{Rising atmospheric methane: 2007-2014 growth and isotopic shift},''
  \emph{Global Biogeochem. Cycles}, vol.~30, no.~9, pp. 1356--1370, 2016. doi:
  10.1002/2016GB005406.

\bibitem{Turner2019}
A.~J. Turner, C.~Frankenberg, and E.~A. Kort, ``{Interpreting contemporary
  trends in atmospheric methane},'' \emph{PNAS}, vol. 116, no.~8, pp.
  2805--2813, 2019. doi: 10.1073/pnas.1814297116.

\bibitem{Bogner2003}
J.~Bogner and E.~Matthews, ``{Global methane emissions from landfills: New
  methodology and annual estimates 1980-1996},'' \emph{Global Biogeochem.
  Cycles}, vol.~17, no.~2, 2003. doi: 10.1029/2002GB001913.

\bibitem{Beauchemin2005}
K.~A. Beauchemin and S.~M. McGinn, ``{Methane emissions from feedlot cattle fed
  barley or corn diets},'' \emph{Journal of Animal Science}, vol.~83, no.~3,
  pp. 653--661, Mar. 2005. doi: 10.2527/2005.833653x.

\bibitem{Allen2013}
D.~T. Allen \emph{et~al.}, ``{Measurements of methane emissions at natural gas
  production sites in the United States},'' \emph{PNAS}, vol. 110, no.~44, pp.
  17\,768--17\,773, Oct. 2013. doi: 10.1073/pnas.1304880110.

\bibitem{Jacob2016}
D.~J. Jacob \emph{et~al.}, ``{Satellite observations of atmospheric methane and
  their value for quantifying methane emissions},'' \emph{Atmospheric Chemistry
  and Physics}, vol.~16, no.~22, pp. 14\,371--14\,396, Nov. 2016. doi:
  10.5194/acp-16-14371-2016.

\bibitem{Duren2016a}
R.~M. Duren \emph{et~al.}, ``{Understanding anthropogenic methane and carbon
  dioxide point source emissions},'' National Academies, RFI-2 White paper for
  the 2017-2027 Decadal Survey for Earth Science and Applications from Space,
  2016.

\bibitem{Frankenberg2011}
C.~Frankenberg \emph{et~al.}, ``{Global column-averaged methane mixing ratios
  from 2003 to 2009 as derived from SCIAMACHY: Trends and variability},''
  \emph{Journal of Geophysical Research: Atmospheres}, vol. 116, no.~D4, p.
  D04302, 2011. doi: 10.1029/2010JD014849.

\bibitem{Strow2003}
L.~Strow, S.~Hannon, S.~{De Souza-Machado}, H.~Motteler, and D.~Tobin, ``{An
  overview of the AIRS radiative transfer model},'' \emph{{IEEE} Trans. Geosci.
  Remote Sens.}, vol.~41, no.~2, pp. 303--313, Feb. 2003. doi:
  10.1109/TGRS.2002.808244.

\bibitem{Yokota2009}
T.~Yokota \emph{et~al.}, ``{Global Concentrations of CO\textsubscript{2} and
  CH\textsubscript{4} Retrieved from GOSAT: First Preliminary Results},''
  \emph{Sola}, vol.~5, pp. 160--163, 2009. doi: 10.2151/sola.2009-041.

\bibitem{Hu2018a}
H.~Hu \emph{et~al.}, ``{Toward Global Mapping of Methane With TROPOMI: First
  Results and Intersatellite Comparison to GOSAT},'' \emph{Geophysical Research
  Letters}, vol.~45, no.~8, pp. 3682--3689, 2018. doi: 10.1002/2018GL077259.

\bibitem{Ayasse2018}
A.~K. Ayasse \emph{et~al.}, ``{Evaluating the effects of surface properties on
  methane retrievals using a synthetic airborne visible/infrared imaging
  spectrometer next generation (AVIRIS-NG) image},'' \emph{Remote Sensing of
  Environment}, vol. 215, pp. 386--397, Jun. 2018. doi:
  10.1016/j.rse.2018.06.018.

\bibitem{Thorpe2014}
A.~K. Thorpe, C.~Frankenberg, and D.~A. Roberts, ``{Retrieval techniques for
  airborne imaging of methane concentrations using high spatial and moderate
  spectral resolution: application to AVIRIS},'' \emph{Atmospheric Measurement
  Techniques}, vol.~7, no.~2, pp. 491--506, Feb. 2014. doi:
  10.5194/amt-7-491-2014.

\bibitem{Thorpe2017}
A.~K. Thorpe \emph{et~al.}, ``{Airborne DOAS retrievals of methane, carbon
  dioxide, and water vapor concentrations at high spatial resolution:
  application to AVIRIS-NG},'' \emph{Atmospheric Measurement Techniques},
  vol.~10, no.~10, pp. 3833--3850, Oct. 2017. doi: 10.5194/amt-10-3833-2017.

\bibitem{Thorpe2013}
A.~K. Thorpe, D.~A. Roberts, E.~S. Bradley, C.~C. Funk, P.~E. Dennison, and
  I.~Leifer, ``{High resolution mapping of methane emissions from marine and
  terrestrial sources using a Cluster-Tuned Matched Filter technique and
  imaging spectrometry},'' \emph{Remote Sensing of Environment}, vol. 134, pp.
  305--318, Jul. 2013. doi: 10.1016/j.rse.2013.03.018.

\bibitem{Thompson2015}
D.~R. Thompson \emph{et~al.}, ``{Real-time remote detection and measurement for
  airborne imaging spectroscopy: a case study with methane},''
  \emph{Atmospheric Measurement Techniques}, vol.~8, no.~10, pp. 4383--4397,
  Oct. 2015. doi: 10.5194/amt-8-4383-2015.

\bibitem{Frankenberg2016}
C.~Frankenberg \emph{et~al.}, ``{Airborne methane remote measurements reveal
  heavy-tail flux distribution in Four Corners region},'' \emph{PNAS}, vol.
  113, no.~35, pp. 9734--9739, Aug. 2016. doi: 10.1073/pnas.1605617113.

\bibitem{Funk2001}
C.~Funk, J.~Theiler, D.~Roberts, and C.~Borel, ``{Clustering to improve matched
  filter detection of weak gas plumes in hyperspectral thermal imagery},''
  \emph{{IEEE} Trans. Geosci. Remote Sens.}, vol.~39, no.~7, pp. 1410--1420,
  Jul. 2001. doi: 10.1109/36.934073.

\bibitem{Theiler2006b}
J.~Theiler and B.~Foy, ``{Effect of Signal Contamination in Matched-Filter
  Detection of the Signal on a Cluttered Background},'' \emph{{IEEE} Geosci.
  Remote Sens. Lett.}, vol.~3, no.~1, pp. 98--102, Jan. 2006. doi:
  10.1109/LGRS.2005.857619.

\bibitem{Manolakis2007}
D.~Manolakis, R.~Lockwood, T.~Cooley, and J.~Jacobson, ``{Robust Matched
  Filters for Target Detection in Hyperspectral Imaging Data},'' in \emph{2007
  IEEE International Conference on Acoustics, Speech and Signal Processing -
  ICASSP '07}.\hskip 1em plus 0.5em minus 0.4em\relax IEEE, 2007. doi:
  10.1109/ICASSP.2007.366733.

\bibitem{Bradley2011}
E.~S. Bradley, I.~Leifer, D.~A. Roberts, P.~E. Dennison, and L.~Washburn,
  ``{Detection of marine methane emissions with AVIRIS band ratios},''
  \emph{Geophysical Research Letters}, vol.~38, no.~10, 2011. doi:
  10.1029/2011GL046729.

\bibitem{Frankenberg2005}
C.~Frankenberg, U.~Platt, and T.~Wagner, ``{Iterative maximum a posteriori
  (IMAP)-DOAS for retrieval of strongly absorbing trace gases: Model studies
  for CH$_4$ and CO$_2$ retrieval from near infrared spectra of SCIAMACHY
  onboard ENVISAT},'' \emph{Atmospheric Chemistry and Physics}, vol.~5, no.~1,
  pp. 9--22, Jan. 2005. doi: 10.5194/acp-5-9-2005.

\bibitem{Manolakis2009b}
D.~Manolakis, R.~Lockwood, T.~Cooley, and J.~Jacobson, ``{Hyperspectral
  detection algorithms: use covariances or subspaces?}'' in \emph{Proc. of
  SPIE}, S.~S. Shen and P.~E. Lewis, Eds., 2009. doi: 10.1117/12.828397.

\bibitem{Donoho2006}
D.~Donoho, ``{Compressed sensing},'' \emph{IEEE Trans. Inform. Theory},
  vol.~52, no.~4, pp. 1289--1306, 2006. doi: 10.1109/TIT.2006.871582.

\bibitem{Lustig2007}
M.~Lustig, D.~Donoho, and J.~M. Pauly, ``{Sparse MRI: The application of
  compressed sensing for rapid MR imaging},'' \emph{Magnetic Resonance in
  Medicine}, vol.~58, no.~6, pp. 1182--1195, 2007. doi: 10.1002/mrm.21391.

\bibitem{Oike2013a}
Y.~Oike and A.~{El Gamal}, ``{CMOS Image Sensor With Per-Column
  $\Sigma$$\Delta$ ADC and Programmable Compressed Sensing},'' \emph{IEEE
  Journal of Solid-State Circuits}, vol.~48, no.~1, pp. 318--328, Jan. 2013.
  doi: 10.1109/JSSC.2012.2214851.

\bibitem{Wiaux2009a}
Y.~Wiaux, L.~Jacques, G.~Puy, A.~M.~M. Scaife, and P.~Vandergheynst,
  ``{Compressed sensing imaging techniques for radio interferometry},''
  \emph{Monthly Notices of the Royal Astronomical Society}, vol. 395, no.~3,
  pp. 1733--1742, May 2009. doi: 10.1111/j.1365-2966.2009.14665.x.

\bibitem{Stevens2015b}
A.~Stevens, L.~Kovarik, P.~Abellan, X.~Yuan, L.~Carin, and N.~D. Browning,
  ``{Applying compressive sensing to TEM video: a substantial frame rate
  increase on any camera},'' \emph{Adv. Structural and Chem. Imaging}, vol.~1,
  no.~1, p.~10, Aug. 2015. doi: 10.1186/s40679-015-0009-3.

\bibitem{Candes2006}
E.~Cand{\`{e}}s, ``{Compressive sampling},'' in \emph{Proceedings of the
  International Congress of Mathematicians Madrid, August 22-30, 2006}.\hskip
  1em plus 0.5em minus 0.4em\relax Zuerich, Switzerland: European Mathematical
  Society Publishing House, 2006, pp. 1433--1452, doi: 10.4171/022-3/69.

\bibitem{Santosa1986}
F.~Santosa and W.~W. Symes, ``{Linear Inversion of Band-Limited Reflection
  Seismograms},'' \emph{SIAM J. on Sci. and Stat. Comput.}, vol.~7, no.~4, pp.
  1307--1330, Oct. 1986. doi: 10.1137/0907087.

\bibitem{Candes2007}
E.~Cand{\`{e}}s and J.~Romberg, ``{Sparsity and incoherence in compressive
  sampling},'' \emph{Inverse Problems}, vol.~23, no.~3, pp. 969--985, Jun.
  2007. doi: 10.1088/0266-5611/23/3/008.

\bibitem{Optimization}
S.~P. Boyd and L.~Vandenberghe, \emph{{Convex Optimization}}.\hskip 1em plus
  0.5em minus 0.4em\relax Cambridge University Press, 2004.

\bibitem{Candes2008a}
E.~J. Cand{\`{e}}s, M.~B. Wakin, and S.~P. Boyd, ``{Enhancing Sparsity by
  Reweighted $\ell$1 Minimization},'' \emph{J. Fourier Anal. Appl.}, vol.~14,
  no.~5, pp. 877--905, 2008. doi: 10.1007/s00041-008-9045-x.

\bibitem{Beck2009}
A.~Beck and M.~Teboulle, ``{A Fast Iterative Shrinkage-Thresholding Algorithm
  for Linear Inverse Problems},'' \emph{SIAM J. Imaging Sci.}, vol.~2, no.~1,
  pp. 183--202, Jan. 2009. doi: 10.1137/080716542.

\bibitem{Manolakis2003}
D.~Manolakis, ``{Detection algorithms for hyperspectral imaging applications: a
  signal processing perspective},'' in \emph{IEEE Workshop on Advances in
  Techniques for Analysis of Remotely Sensed Data, 2003}, 2003. doi:
  10.1109/WARSD.2003.1295218.

\bibitem{Kraut2005b}
S.~Kraut, L.~Scharf, and R.~Butler, ``{The adaptive coherence estimator: a
  uniformly most-powerful-invariant adaptive detection statistic},'' \emph{IEEE
  Transactions on Signal Processing}, vol.~53, no.~2, pp. 427--438, Feb. 2005.
  doi: 10.1109/TSP.2004.840823.

\bibitem{Berk2014}
A.~Berk, P.~Conforti, R.~Kennett, T.~Perkins, F.~Hawes, and J.~van~den Bosch,
  ``{MODTRAN{\textregistered} 6: A major upgrade of the
  MODTRAN{\textregistered} radiative transfer code},'' in \emph{2014 6th
  Workshop on Hyperspectral Image and Signal Processing: Evolution in Remote
  Sensing (WHISPERS)}, 2014. doi: 10.1109/WHISPERS.2014.8077573.

\bibitem{Theiler2012}
J.~Theiler, ``{The incredible shrinking covariance estimator},'' in \emph{Proc.
  of SPIE}, 2012. doi: 10.1117/12.918718.

\bibitem{Thompson2016}
D.~R. Thompson \emph{et~al.}, ``{Space-based remote imaging spectroscopy of the
  Aliso Canyon CH$_4$ superemitter},'' \emph{Geophysical Research Letters},
  vol.~43, no.~12, pp. 6571--6578, 2016. doi: 10.1002/2016GL069079.

\bibitem{Hamlin2011}
L.~Hamlin \emph{et~al.}, ``{Imaging spectrometer science measurements for
  Terrestrial Ecology: AVIRIS and new developments},'' in \emph{2011 Aerospace
  Conference}, 2011. doi: 10.1109/AERO.2011.5747395.

\bibitem{Thompson2018}
D.~R. Thompson, V.~Natraj, R.~O. Green, M.~C. Helmlinger, B.-C. Gao, and M.~L.
  Eastwood, ``{Optimal estimation for imaging spectrometer atmospheric
  correction},'' \emph{Remote Sensing of Environment}, vol. 216, pp. 355--373,
  Oct. 2018. doi: 10.1016/j.rse.2018.07.003.

\bibitem{Dennison2013a}
P.~E. Dennison \emph{et~al.}, ``{High spatial resolution mapping of elevated
  atmospheric carbon dioxide using airborne imaging spectroscopy: Radiative
  transfer modeling and power plant plume detection},'' \emph{Remote Sensing of
  Environment}, vol. 139, pp. 116--129, Dec. 2013. doi:
  10.1016/j.rse.2013.08.001.

\bibitem{Thompson2015a}
D.~R. Thompson, B.-C. Gao, R.~O. Green, D.~A. Roberts, P.~E. Dennison, and
  S.~R. Lundeen, ``{Atmospheric correction for global mapping spectroscopy:
  ATREM advances for the HyspIRI preparatory campaign},'' \emph{Remote Sensing
  of Environment}, vol. 167, pp. 64--77, Sep. 2015. doi:
  10.1016/j.rse.2015.02.010.

\bibitem{Savitzky1964}
A.~Savitzky and M.~J.~E. Golay, ``{Smoothing and Differentiation of Data by
  Simplified Least Squares Procedures.}'' \emph{Analytical Chemistry}, vol.~36,
  no.~8, pp. 1627--1639, Jul. 1964. doi: 10.1021/ac60214a047.

\bibitem{Schlapfer2011}
D.~Schl{\"{a}}pfer and R.~Richter, ``{Spectral polishing of high resolution
  imaging spectroscopy data},'' in \emph{7th SIG-IS Workshop on Imaging
  Spectroscopy}.\hskip 1em plus 0.5em minus 0.4em\relax Edinburgh: Edinburgh
  Publication, 2011.

\bibitem{Nagler2000}
P.~Nagler, C.~Daughtry, and S.~Goward, ``{Plant Litter and Soil Reflectance},''
  \emph{Remote Sensing of Environment}, vol.~71, no.~2, pp. 207--215, Feb.
  2000. doi: 10.1016/S0034-4257(99)00082-6.

\bibitem{Jongaramrungruang2018}
S.~Jongaramrungruang \emph{et~al.}, ``Towards accurate methane point-source
  quantification from high-resolution 2-d plume imagery,'' \emph{Atmospheric
  Measurement Techniques}, vol.~12, no.~12, pp. 6667--6681, 2019. doi:
  10.5194/amt-12-6667-2019.

\bibitem{NIPS2019_9015}
A.~Paszke \emph{et~al.}, ``Pytorch: An imperative style, high-performance deep
  learning library,'' in \emph{Advances in Neural Information Processing
  Systems 32}, H.~Wallach, H.~Larochelle, A.~Beygelzimer, F.~d\textquotesingle
  Alch\'{e}-Buc, E.~Fox, and R.~Garnett, Eds.\hskip 1em plus 0.5em minus
  0.4em\relax Curran Associates, Inc., 2019, pp. 8024--8035.

\bibitem{Dennison2013b}
P.~E. Dennison, A.~K. Thorpe, D.~A. Roberts, and R.~O. Green, ``{Modeling
  sensitivity of imaging spectrometer data to carbon dioxide and methane
  plumes},'' in \emph{2013 5th Workshop on Hyperspectral Image and Signal
  Processing: Evolution in Remote Sensing (WHISPERS)}, 2013. doi:
  10.1109/WHISPERS.2013.8080614.

\bibitem{Theiler2006a}
J.~Theiler, B.~R. Foy, and A.~M. Fraser, ``{Nonlinear signal contamination
  effects for gaseous plume detection in hyperspectral imagery},'' in
  \emph{Proc. SPIE 6233, Algorithms and Technologies for Multispectral,
  Hyperspectral, and Ultraspectral Imagery XII}, 2006. doi: 10.1117/12.665608.

\end{thebibliography}
\begin{IEEEbiography}[{\includegraphics[width=1in,height=1.25in,clip,keepaspectratio]{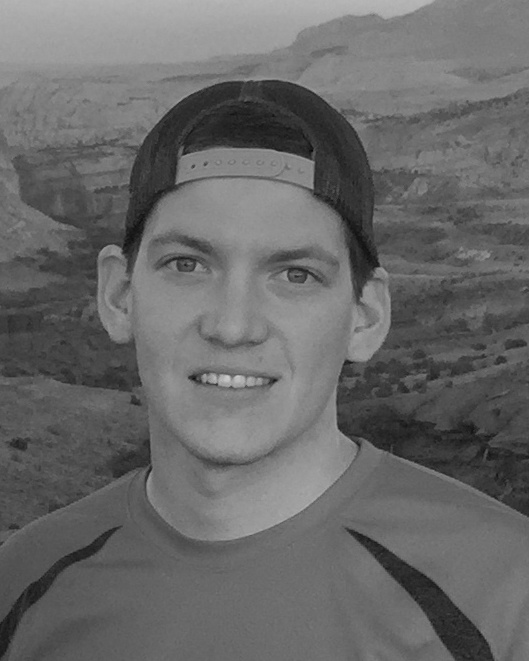}}]{Markus D. Foote}
(GS'20) received the B.Sc. degree in biomedical engineering from the University of Utah, Salt Lake City, UT, USA, in 2015 and the M.Sc. degree in biomedical engineering from the University of Utah in 2019. He is currently pursuing the Ph.D. degree in biomedical engineering from the University of Utah. He is a member of the Scientific Computing and Imaging Institute. 

His research interests include machine learning, image processing and analysis, computational anatomy, and accelerated computing.

\end{IEEEbiography}
\begin{IEEEbiography}[{\includegraphics[width=1in,height=1.25in,clip,keepaspectratio]{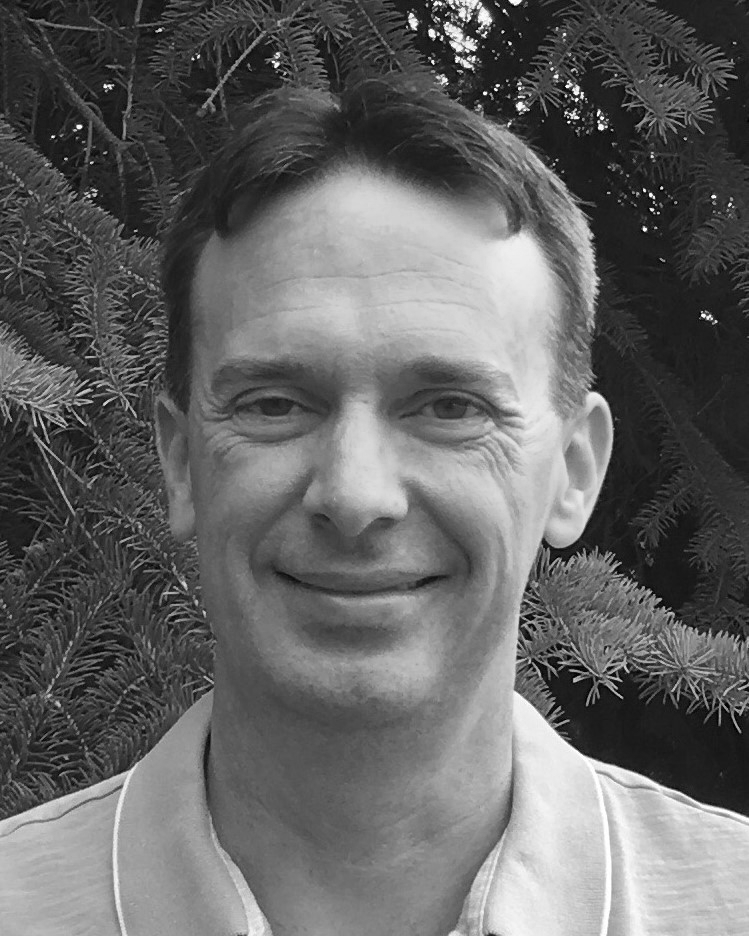}}]{Philip E. Dennison}
received the M.A. degree and Ph.D. degree from the Department of Geography, University of California Santa Barbara, in 1999 and 2003, respectively.

He is currently professor and chair of the Department of Geography at the University of Utah. He is also the director of the Utah Remote Sensing Applications Lab. His research interests include imaging spectroscopy, remote sensing of vegetation, atmospheric trace gas retrievals, wildfire, and firefighter safety.

\end{IEEEbiography}
\begin{IEEEbiography}[{\includegraphics[width=1in,height=1.25in,clip,keepaspectratio]{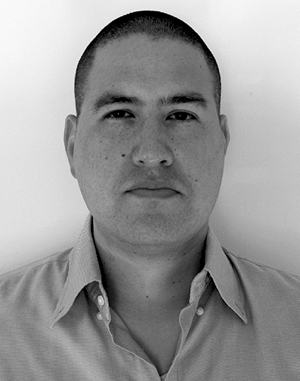}}]{Andrew K. Thorpe}
received the B.Sc. degree in geological sciences from Brown University, Providence, RI, USA, in 2004 and the Ph.D. degree in geography with a remote sensing focus from the University of California, Santa Barbara, Santa Barbara, CA, USA, in 2015. 

He is a research technologist in the imaging spectroscopy group at the Jet Propulsion Laboratory, California Institute of Technology, Pasadena, CA, USA. For the last decade, Dr. Thorpe has worked with imaging spectrometer data and developed methods for high spatial resolution mapping of methane and carbon dioxide point source emissions using imaging spectrometers like AVIRIS and AVIRIS-NG. He has developed quantitative gas retrievals using differential optical absorption spectroscopy, methods to estimate emission rates, and contributed to projects aimed at visualizing methane results and reducing data latency. He is also involved in ongoing efforts to develop instrument concepts for dedicated imaging spectrometers for trace gas mapping and quantification.

Dr. Thorpe is a member of the American and European Geophysical Unions and participates as a reviewer for a number of academic journals. He was a recipient of a NASA Earth and Space Science Fellowship (NESSF), a JPL Voyager and Discovery Award, and a number of additional JPL Team and Group Achievement Awards. 

\end{IEEEbiography}
\begin{IEEEbiography}[{\includegraphics[width=1in,height=1.25in,clip,keepaspectratio]{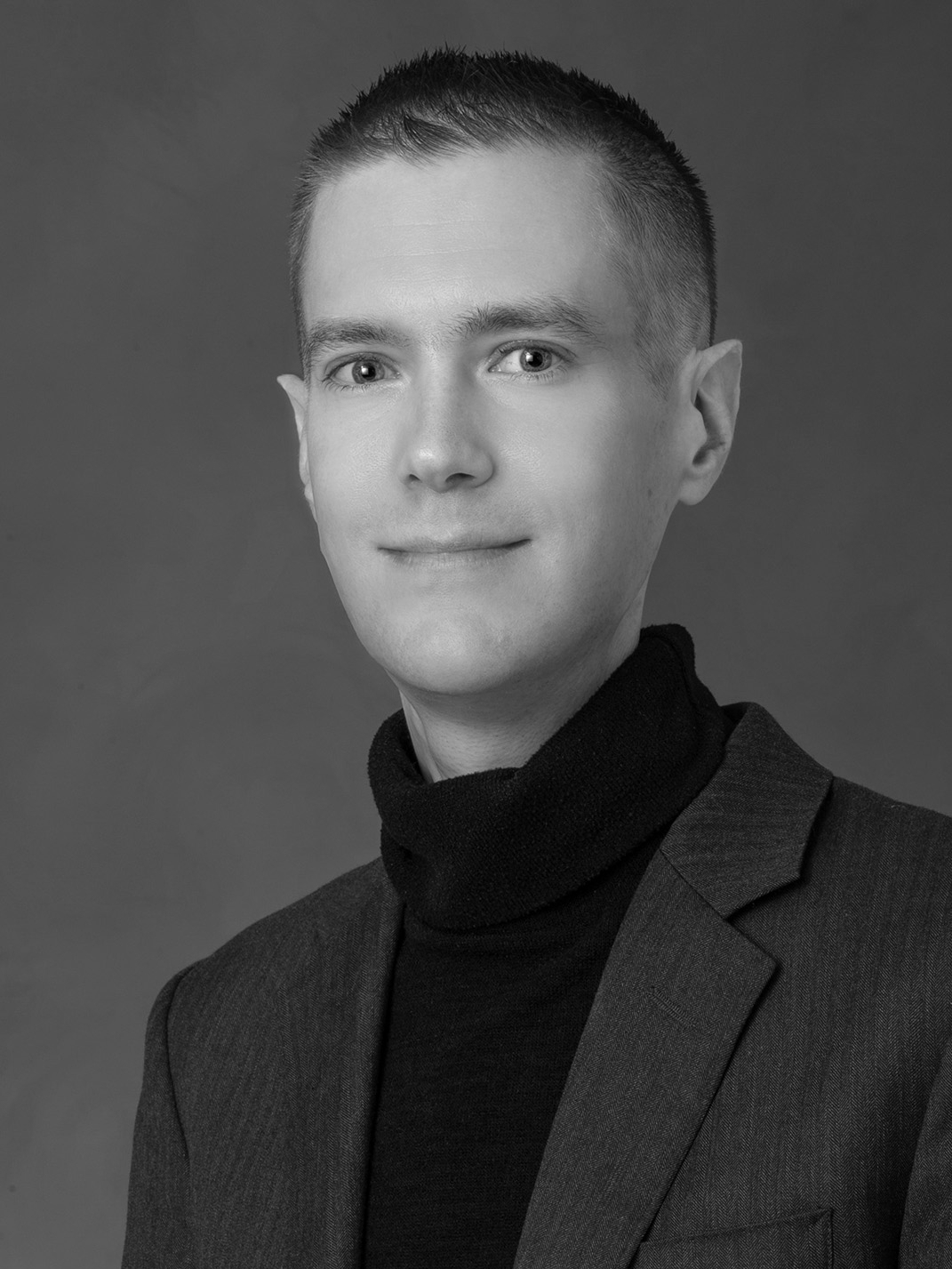}}]{David R. Thompson}
received the B.A.
degree in computer science from the Carleton
College, Northfield, MN, USA, the M.Sc. degree
in informatics from the University of Edinburgh,
Edinburgh, U.K., and the Ph.D. degree in robotics
from the Robotics Institute, Carnegie Mellon University, Pittsburgh, PA, USA.

He is currently a Technical Group Lead of the JPL Imaging Spectroscopy Group, Instrument Scientist for NASA's EMIT mission and Lunar Trailblazer, and Investigation Scientist for NASA's Airborne Visible Infrared Imaging Spectrometer (AVIRIS) project.  His research advances the algorithms and practice of imaging spectroscopy for characterizing Earth and other planetary bodies.  He has published over 80 papers in areas of spectroscopy, remote sensing, and data science for remote exploration.
 
\end{IEEEbiography}
\vfill
\newpage
\begin{IEEEbiography}[{\includegraphics[width=1in,height=1.25in,clip,keepaspectratio]{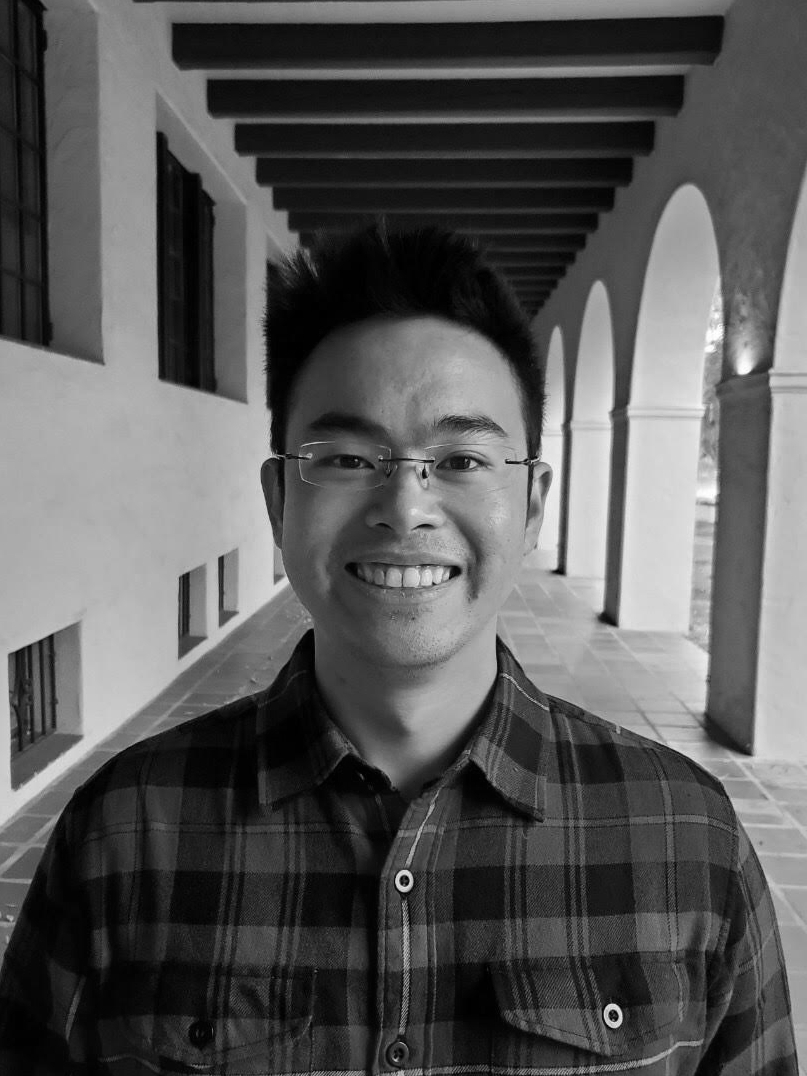}}]{Siraput Jongaramrungruang}
received the B.A. and M.Sc. degrees in physics from the University of Cambridge, Cambridge, UK, 
the M.Sc. degree in environmental science and engineering from the California Institute of Technology, Pasadena, CA, USA, 
and is currently a Ph.D. candidate in environmental science and engineering from the California Institute of Technology.

Mr. Jongaramrungruang is a recipient of a NASA Earth and Space Science Fellowship (NESSF).
\end{IEEEbiography}
\begin{IEEEbiography}[{\includegraphics[width=1in,height=1.25in,clip,keepaspectratio]{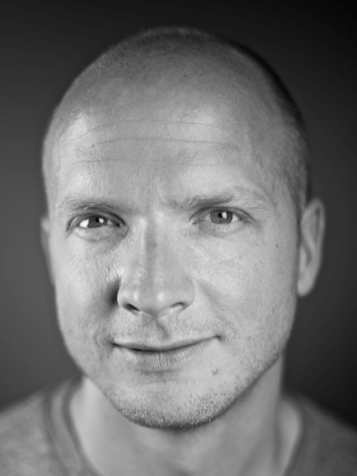}}]{Christian Frankenberg}
received his Ph.D. degree from the Institute of Environmental Physics,
Heidelberg, Germany, in 2005, studying atmospheric
methane and carbon monoxide from the
SCIAMACHY satellite.

After his postdoctoral fellowship with the
Netherlands Institute for Space Research, Utrecht,
The Netherlands (2006-2009), he
has been with the Jet Propulsion Laboratory,
California Institute of Technology (Caltech),
Pasadena, CA, USA, as a Scientist since January 2010. As of September
2015, he is a Professor for environmental science and engineering
with Caltech. He is currently working on greenhouse gas and fluorescence
retrievals from space, ground and air as well as global land surface modeling.

\end{IEEEbiography}
\begin{IEEEbiography}[{\includegraphics[width=1in,height=1.25in,clip,keepaspectratio]{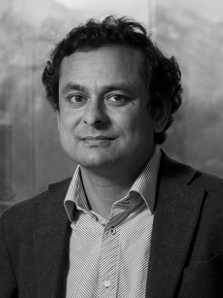}}]{Sarang C. Joshi}
received the D.Sc. degree in electrical engineering from Washington University in St. Louis, St. Louis, MO, USA, in 1998. He is currently professor of Biomedical Engineering at University of Utah, Salt Lake City, UT, USA, and adjunct professor in Departments of Computer Science, Mathematics and Radiation Oncology. He is member of the Scientific Computing and Imaging Institute. 

His research interests include mathematical image analysis, computer vision and computational anatomy. He has won numerous awards including the David Marr Best Paper Award, The international journal Signal Processing Most cited paper Award, and MICCAI Best of the Journal Issue Award.
\end{IEEEbiography}
\vfill
\end{document}